\documentclass[sigconf]{acmart}
\usepackage{ragged2e} 
\usepackage{algpseudocode}
\usepackage{booktabs,makecell, multirow, tabularx}
\usepackage{dsfont}
\usepackage{bm}

\usepackage{enumitem}
\usepackage{verbatim}
\usepackage[ruled,vlined]{algorithm2e}
\usepackage{multirow}
\usepackage{subfigure} 
\usepackage{ragged2e}
\usepackage{caption}
\usepackage{graphicx}
\usepackage[normalem]{ulem}
\useunder{\uline}{\ul}{}
\usepackage{amsmath}
\usepackage{setspace}
\usepackage{bm}

\newcommand{\ie}{\emph{i.e., }}
\newcommand{\eg}{\emph{e.g., }}

\newcommand{\wrt}{\emph{w.r.t. }}

\AtBeginDocument{%
  }

\setcopyright{none}
\renewcommand\footnotetextcopyrightpermission[1]{}
\begin{document}

\title{Precision Profile Pollution Attack on Sequential Recommenders via Influence Function}

\author{Xiaoyu Du}
\email{duxy@njust.edu.cn}
% \orcid{0009-0009-1747-5962}
\authornote{Xiaoyu Du and Yingying Chen are co-first authors.}
\affiliation{%
  \institution{Nanjing University of Science and Technology}
  \city{Nanjing}
  \country{China}
}

\author{Yingying Chen}
\email{YingyingChen@njust.edu.cn}
\orcid{0009-0009-1747-5962}
\authornotemark[1]

\affiliation{%
  \institution{Nanjing University of Science and Technology}
  \city{Nanjing}
  \country{China}
}

% \author{Xiaoyu Du}
% \affiliation{%
%   \institution{Nanjing University of Science and Technology}
%   \city{Nanjing}
% \email{duxy@njust.edu.cn}

%  \author{Yingying Chen}
%  \authornotemark[1]
% \affiliation{%
%   \institution{Nanjing University of Science and Technology}
%   \city{Nanjing}
% \email{YingyingChen@njust.edu.cn}

\author{Yang Zhang}
\affiliation{%
  \institution{University of Science and Technology of China}
  \city{Hefei}
  \country{China}
}
\email{zy2015@mail.ustc.edu.cn}

\author{Jinhui Tang}
\authornote{Jinhui Tang is the corresponding author.}
\affiliation{%
 \institution{Nanjing University of Science and Technology}
 \city{Nanjing}
 \country{China}}
\email{jinhuitang@njust.edu.cn}

% \renewcommand{\shortauthors}{Du et al.}

\begin{abstract}
  Sequential recommendation approaches have demonstrated remarkable proficiency in modeling user preferences. Nevertheless, they are susceptible to profile pollution attacks (PPA), wherein items are introduced into a user's interaction history deliberately to influence the recommendation list. Since retraining the model for each polluted item is time-consuming, recent PPAs estimate item influence based on gradient directions to identify the most effective attack candidates. However, the actual item representations diverge significantly from the gradients, resulting in disparate outcomes.
To tackle this challenge, we introduce an \textbf{INF}luence Function-based \textbf{Attack} approach~(INFAttack) that offers a more accurate estimation of the influence of polluting items. Specifically, we calculate the modifications to the original model using the influence function when generating polluted sequences by introducing specific items. Subsequently, we choose the sequence that has been most significantly influenced to substitute the original sequence, thus promoting the target item. Comprehensive experiments conducted on five real-world datasets illustrate that INFAttack surpasses all baseline methods and consistently delivers stable attack performance for both popular and unpopular items.
\end{abstract}





\maketitle

\section{Introduction}

Sequential Recommendation (SR)~\cite{yu2023self} aims to provide personalized recommendations by predicting user preferences through modeling the temporal sequence of interactions between users and items~\cite{DBLP:conf/cikm/LuoHLP022,DBLP:journals/tkdd/ShenOL22,zhou2020s3}. 
Commonly, SR methods employ sequence modeling tools like Markov chains (MC)~\cite{DBLP:conf/icdm/HeM16}, convolutional neural networks (CNN)~\cite{DBLP:conf/wsdm/TangW18}, and recurrent neural networks (RNN)~\cite{DBLP:journals/corr/HidasiKBT15} to describe user historical interactions. Recently, SR models have also utilized Transformers ~\cite{DBLP:conf/nips/VaswaniSPUJGKP17,DBLP:conf/naacl/DevlinCLT19} as sequence encoders to capture item relevance and obtain high-quality sequence representations ~\cite{DBLP:conf/icdm/KangM18,yuan2020parameter}. Because of their ability to accurately model user representations using interaction information, SR has gained widespread attention and shown excellent performance in recommendation tasks~\cite{DBLP:conf/cikm/ZhangYG0QG0LT22,DBLP:conf/cikm/JiangZLLKZW0023,zhou2022filter}.

However, SR is vulnerable to attacks, which can easily manipulate the system to produce desired results by attackers~\cite{DBLP:conf/ndss/YangGC17,DBLP:journals/tifs/ZhangXHWZJ20}. One common attack is profile pollution attack, where attackers fabricate or manipulate user behavior data to interfere with the recommendation system's results~\cite{DBLP:conf/cikm/SunLWPLOJ19}. This attack stems from two basic conclusions: 1) distillation techniques can convert black-box attacks into corresponding white-box attacks, making the proposed white-box attack mode still effective on black-box models~\cite{DBLP:conf/recsys/YueHZM21}; 2) injecting data can falsify user interaction histories in logs, thus contaminating user profiles~\cite{DBLP:conf/kbse/LeeHR17}. profile pollution attack has thus become an important branch of sequential recommendation attacks ~\cite{meng2014your}, with the key aspect being the selection of appropriate injection samples to achieve the target item exposure rate~\cite{DBLP:conf/sp/Carlini017}. However, the impact of injection samples on the model can only be determined through model retraining, which can be resource and time-consuming.

\begin{figure}[htp]
    \centering
    \includegraphics[width=0.7\linewidth]{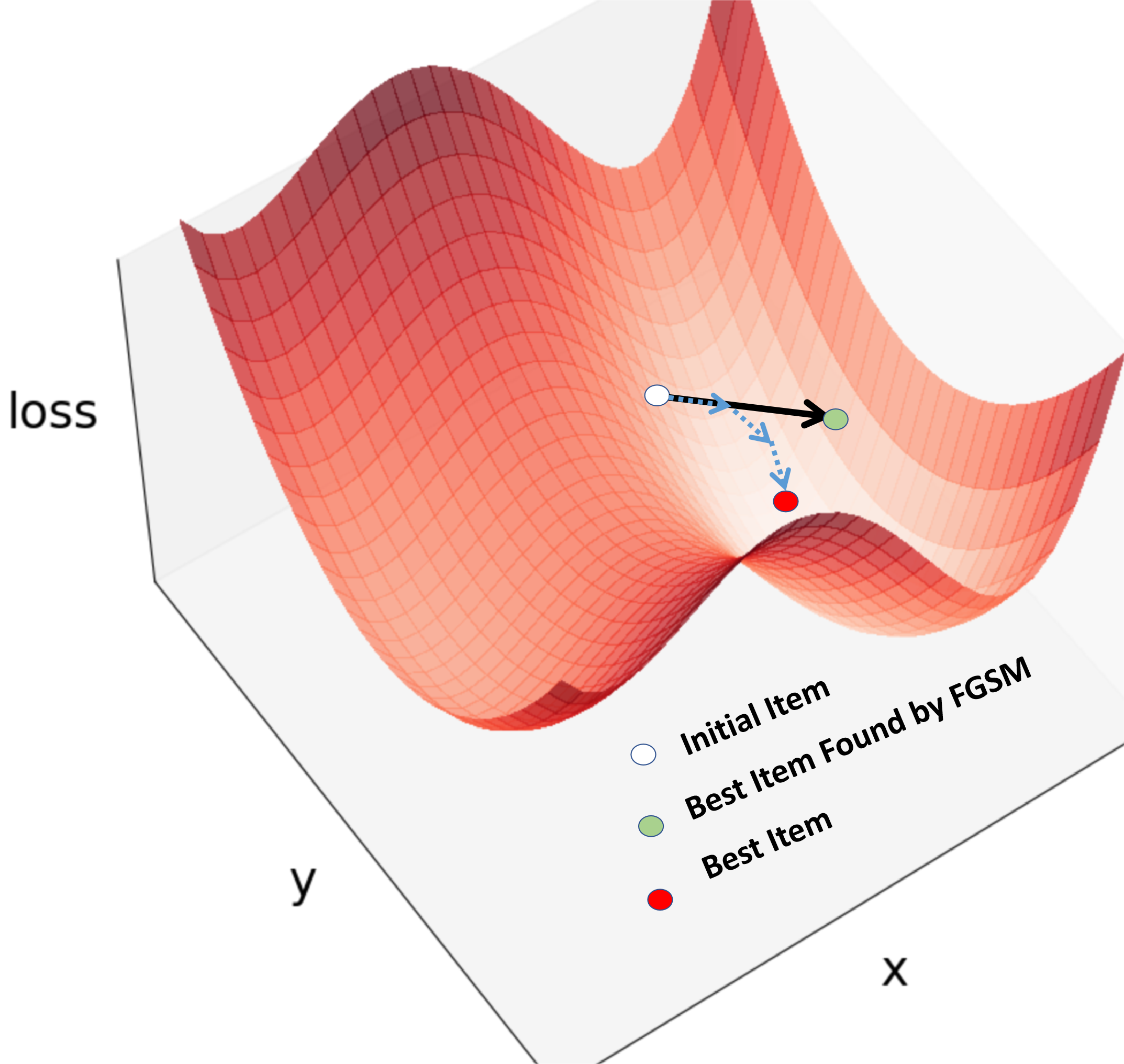}
    \caption{Disadvantages of the FGSM attack method. The white circle, green circle, and red circle indicate the initial item, the optimal item found by the current gradient method, and the real optimal item, respectively.}
     % \vspace{-3mm}
    \label{fig:gradient}
    \vspace{-3mm}
\end{figure}

% Recent attack strategies have relied on gradient estimation to evaluate the impact of samples on results~\cite{DBLP:conf/recsys/YueHZM21}. They create perturbations based on the gradient of the attack function and inject them into user interaction sequences to promote target items. 
% Similarly, \citet{DBLP:conf/recsys/YueZKSW22} employ this method to generate adversarial items that replace original items in the sequence to promote the target item. 
Recent attack strategies have relied on gradient estimation to evaluate the impact of samples on results~\cite{DBLP:conf/recsys/YueHZM21,DBLP:conf/recsys/YueZKSW22}. 
% They create perturbations based on the gradient of the attack function and inject them into user interaction sequences to promote target items. 
They create perturbations based on the gradient of the attack function and inject them into user interaction sequences to promote target items.
% Similarly, \citet{DBLP:conf/recsys/YueZKSW22} employ this method to generate adversarial items that replace original items in the sequence to promote the target item. 
Although these methods are effective, they suffer from three shortcomings. 
Firstly, to construct attack data, they leverage only a single-step gradient computation, as shown by the black arrow in Figure~\ref{fig:gradient}. However, finding the optimal attack samples typically requires multiple iterations of gradient computation in theory~\cite{DBLP:conf/iclr/MadryMSTV18}, as shown by the blue arrow in Figure~\ref{fig:gradient}. This insufficient computation results in suboptimal attack sample construction. Secondly, ideal attack samples generated based on gradients may not exist in the dataset, and the difference between similar and ideal samples can further impact the effectiveness of the attack. 
Lastly, such gradient-based attack strategies only use gradients to affect parameter changes but cannot guarantee the model's effectiveness in recommending target items after model training, and consequently, cannot ensure the effectiveness of the modified sequence in recommending target items.
% Lastly, these gradient-based strategies fail to account for the impact of retraining when constructing attack data, potentially constraining their effectiveness in practical attacks conducted through retraining.

% Lastly, such gradient-based strategies fail to measure the true influence of the injection data in recommending target items after retraining, limiting their real attacking effectiveness.

% Firstly, they utilize single-step gradient computation to roughly estimate the sample direction and search for relevant samples in the gradient direction (the black arrow in Figure (\ref{fig:gradient})), resulting in the green circle sample. However, in practice, gradient attacks typically require multiple iterations (the blue arrow in the figure) to find the optimal sample, as demonstrated in theoretical studies ~\cite{DBLP:conf/iclr/MadryMSTV18}.
% \ref{sample/imag/introduction.png}
\graphicspath{{img/}}

To address the aforementioned issues, we propose a method that can estimate the impact of modifications made to training samples on model parameters. 
This ensures that the effect of attack samples on model predictions is known and effective. 
Inspired by~\cite{DBLP:conf/icml/KohL17}, we present an attack scheme based on influence functions that selects the best attack sample by estimating the effect of contaminated samples on model predictions. The method directly evaluates the final model formed by the pollution term, providing a relatively accurate estimate of the true outcome and achieving high accuracy.

The contributions of this paper are summarized as follows:
\begin{itemize}[leftmargin=*]
    \item We propose an influence-based Profile Pollution attack algorithm, INFAttack. By designing an influence-based injection module, effective attacks on sequential recommendation can be generated.  
    \item In the case of effective attacks, we explored the stability of attacks in different popularity items. Our proposed method maintains effectiveness across different popularity items.
    \item We evaluate our attacks on five real-world datasets. The experimental results show that our attack method is superior to existing attack methods, and also has effective promotion effects on unwanted target items.
\end{itemize}
% • We propose an influence-based file contamination attack algorithm, INFAttack. By designing an influence-based injection module, effective attacks on sequence recommendation can be generated. In addition, compared with existing model-related attacks, our framework can more stably perform file contamination attacks. 

% • We theoretically demonstrate that, subject to an effective attack, we explore an influence-based approach to estimate the impact of a polluted sequence. Through theoretical analysis, the effectiveness of influence function in pollution attack is proved, which provides a new way to ensure the effectiveness of sequence recommendation pollution attack.

% • We evaluate our attacks on three real-world datasets. The experimental results show that our attack method is superior to existing attack methods, and also has effective promotion effects on unwanted target items.
\section{related work}

%%% 建议：
%%% P1: attack on recommender system
%%%     说总体有两类，介绍定义。然后一笔带过data poisoning ，转到profile pollution。
%%%     分类概括profile pollution方法。 讲出这些方法与本文方法的区别。
%%%     然后可以单独把针对sequential的方法领出来说下，讲下区别。

%%% P1: influence function in sequential
%%%     总体介绍下influence function，讲influence function在推荐中的几种应用。
%%%     专门讲influecne function 关于推荐 attack的工作，突出本文与他们的区别。
%%% 参考，不要直接照着写：Influencefunctionis aclassic concept originating from robust statistics[10], which is used to estimate the influence of data on learning models. It has been widely used for various tasks in various areas[9,16,37].In recommendation, it has also been utilized for dealing with various tasks in recent research. First, some efforts[8,30,32,35] apply the influence function to achieve more efficient and effective recommendation attacks. For example,[30] utilizes the influence function to quickly estimate the quality of generated fake users, helping find more valid fake users.Second,the influence function has also been used for debiasing[33] [33] utilizes the influence function to quantify how training points are essential for the model performance on unbiased data and further takes the estimated importancescores to reweight training data for debiasing. Besides,[5]utilizes the influence function to explain the recommendation results,which first measures the influences of training interactions on predictions and then provides neighbor-style explanations based on the most influential training interactions. Different from these researches, we take influence function to achieve recommendation unlearning.

\subsection{Attacks on Recommender Systems}
Recommender systems are vulnerable to various types of recommendation attacks.
According to the attacking strategies, recommendation attacks can be divided into two categories: data poisoning attacks and profile pollution attacks. Data poisoning attacks ~\cite{DBLP:conf/acsac/FangYGL18,DBLP:conf/www/ZhangLD020, DBLP:journals/access/SundarLZGR20,DBLP:conf/ndss/HuangMGL0X21} involve generating fake users by forging user-profiles and injecting them into the training data of the recommender system, thereby resulting in biased recommendations expected by attackers. In contrast, profile pollution attacks ~\cite{DBLP:conf/kbse/LeeHR17,DBLP:conf/uss/XingMDSFL13} are more subtle, and attackers interfere with the recommendation results of recommender systems by modifying, tampering, or forging user behavior data.

In this research, we concentrate on profile pollution and execute user-specific attacks to manipulate the recommendation outcomes~\cite{DBLP:conf/recsys/YueHZM21}. Profile pollution attacks can be classified into four categories based on the attacker's means and intention ~\cite{deldjoo2021survey}.
The first category is the profile injection attack, which involves injecting false information to deceive the recommendation system ~\cite{DBLP:conf/uss/XingMDSFL13,DBLP:conf/ccs/MengXSWL14,DBLP:journals/tifs/ZhangXHWZJ20}. For instance, Wei et al.\cite{DBLP:conf/ccs/MengXSWL14} propose an ad spoofing mechanism to manipulate the ad selection process by injecting false information into users' interest tags. Similarly,~\cite{DBLP:journals/tifs/ZhangXHWZJ20} tampers with the content of HTTP web pages using web injection to force users to access specific items in certain web services.
The second category is the profile replacement attack, where the attacker replaces the partial user profiles (\eg historical interactions) with carefully selected content (\eg items) to carry out the attack~\cite{DBLP:conf/recsys/YueZKSW22}. One representative study is ~\cite{DBLP:conf/recsys/YueZKSW22}, which uses the attack loss gradient to guide the selection of replacement items.
The third category is the repetition and deceptive behavior attack, which is a specific type of profile injection attack that repeatedly injects some specific items such as target items ~\cite{DBLP:conf/wsdm/TangW18}.
The fourth category is the recommendation knowledge-based attack, which utilizes expert knowledge on recommendations to attack recommender systems ~\cite{DBLP:conf/ndss/YangGC17}. For example, ~\cite{DBLP:conf/ndss/YangGC17} performs false co-access injection inspired by the knowledge that collaborative information is crucial for the recommendation.
The attack method proposed in our study falls under the profile injection attack category. The most significant difference between our approach and the above methods is that we do not use gradient, similarity, or recommendation knowledge to perform attacks. Instead, we perform more powerful attacks by leveraging the influence function to directly estimate the influence of injections on promoting the target items.

\subsection{Influence Function in Sequential Recommendation}

Influence function is a classic concept originating from robust statistics ~\cite{lambert1981influence,ruppert1987kurtosis}, which is a powerful tool for measuring the influence of data on a model's prediction results~\cite{DBLP:conf/sp/LecuyerAG0J19}.  It is applicable to many differentiable models and has been widely used to solve various tasks (\eg prediction explanation~\cite{koh2019accuracy}, unlearning~\cite{unlearnign}, denoising~\cite{inf-relabel}) across different fields such as computer vision and natural language processing.

The influence function has also been applied to deal with various tasks in recommendation.
For instance, it has been used to generate explanations for recommendations by estimating the influence of training interactions on predictions and providing the most influential training interactions as explanations~\cite{millecamp2019explain}. Meanwhile, it has been used to remove data bias and noise, leading to more accurate and fairer recommendations ~\cite{wang2022understanding}. Specifically,~\cite{wang2022understanding} uses the influence function to quantify how training points affect the model's performance on unbiased data and reweights training data for debiasing. The influence function has also been used for designing poisoning resistance training algorithms ~\cite{yu2020influence}, which improve the robustness of recommendation systems. Furthermore, the influence function has been used for data poisoning attacks with different goals and utilization methods  ~\cite{wu2021triple,zhang2020practical,fang2020influence}. ~cite{fang2020influence} uses the influence function to select a subset of normal users who are influential to the recommendations and focuses on these normal users to efficiently achieve the attack target. ~\cite{zhang2020practical} proposes an RL-based data poisoning attack method that uses the influence function to generate an influence-based reward reflecting the attacking quality for the RL-based attacker. ~\cite{wu2021triple} proposes to use the influence function to generate fake users to bring more attack revenue.

In contrast to existing works, we utilize the influence function to achieve effective profile pollution attacks, which differ inherently from the tasks they studied. While our method is more related to studies on data poisoning attacks, there are still differences between them, such as the influence of injected items could be more difficult to estimate.

% Different from these studies, we adopt the influence function to implement anti-learning for recommendation, and also different from previous application domains, we select the optimal items to inject according to the influence function in profile pollution attack on sequential recommendation.
\section{Problem Formulation}

\textbf{Sequential Recommendation}. We denote a sequential recommender model by $f_\theta$, where $\theta$ represents the learnable parameters of the model. Let $u\in \mathcal{U}$ and $v\in \mathcal{V}$ denote a user and an item, respectively, where $\mathcal{U}$ ($\mathcal{V}$) denotes the set of all users (items). We assume there are $n$ users and $m$ items, \textit{i.e.}, $|\mathcal{U}|=n$ and $|\mathcal{V}|=m$, and each user associates with a chronological sequence $\bm{x}_{u}=[v_1,\dots,v_t,\dots,v_{T}]$, where $v_{t}$ denotes the $t$-th interacted item by $u$. Let $\mathcal{X}=\{\bm{x}_{u}|u\in\mathcal{U}\}$ denote the interaction sequences of all user. Then, the model parameters of the recommender model $f_\theta$ are learned by fitting $\mathcal{X}$. Formally, the learning objective is defined as
\begin{equation} \label{eq:training-loss}
    \hat{\theta} = argmin_{\theta} \frac{1}{n} \sum_{u\in\mathcal{U}} L(\bm{x}_{u};\theta),
\end{equation}
where $L(\bm{x}_{u},\theta)$ is the loss for fitting $u$'s sequence $\bm{x}_{u}$, which can be varied due to the choice of training strategy of the sequential model\footnote{For example, it could be different for SASRec~\cite{DBLP:conf/icdm/KangM18} and Bert4Rec~\cite{DBLP:conf/cikm/SunLWPLOJ19}. Specifically, we use the loss taken by the paper proposing the recommender model.}. During testing, the learned model $f_{\hat{\theta}}$ could generate a recommendation score for each candidate item, indicating how likely the item would be the $(T\text{+}1)$-th interacted item. The final recommendation list is generated based on the scores.

\begin{figure}
    \centering
    \includegraphics[width=\linewidth]{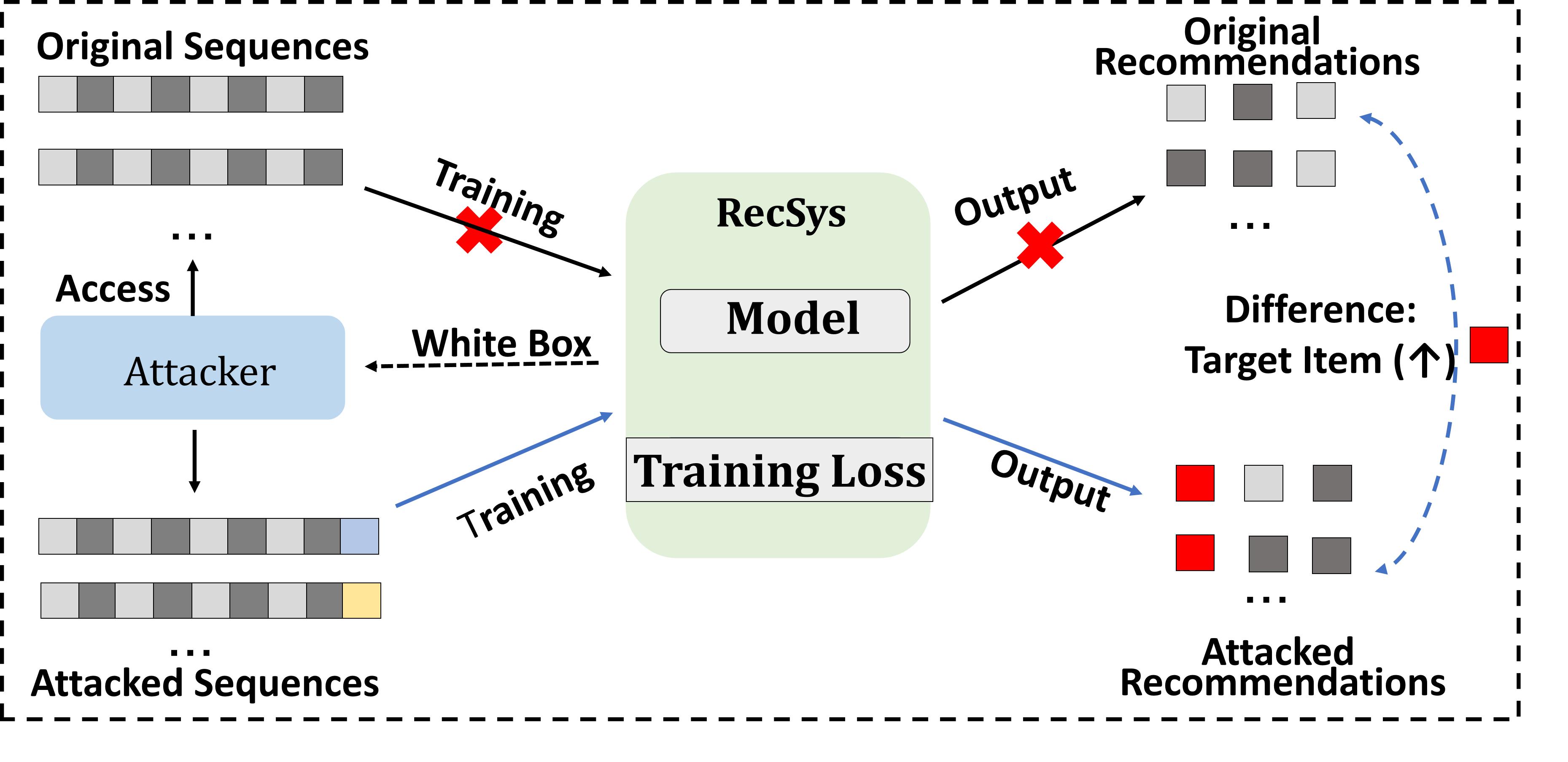}
    \caption{Profile pollution attack problem defined in this work for sequential recommendation:  an attacker injects certain items into a user's interaction sequence to create polluted sequences, such that the model trained on the polluted data could enhance the recommendation of a target item compared to the model trained on the clearing data.}
     \vspace{-2mm}
    \label{fig:attack-flow}
    \vspace{-2mm}
\end{figure}

\vspace{+5pt}
\noindent \textbf{Profile Pollution Attack}. In this work, as shown in Figure~\ref{fig:attack-flow}, we consider the scenario where an attacker can inject certain items into the training sequences of users to perform a profile pollution attack in sequential recommendation, aiming at boosting the recommendation of a target item. Specifically, similar to previous work~\cite{DBLP:conf/recsys/YueZKSW22}, we make the following assumptions to define our research scope:

\begin{itemize}[leftmargin=*]
\item \textbf{Data Access and Manipulation}: The attacker has access to the original training data $\mathcal{X}$ and item set $\mathcal{V}$. For each training sequence $\bm{x}_{u}=[v_{1},\dots,v_{T}] \in \mathcal{X}$, the attacker can insert $K$ items $[v_{1}^{\prime},\dots, v_{K}^{\prime}]$ (\eg via malware~\cite{DBLP:conf/kbse/LeeHR17}) to construct an attacked sequence $\bm{x}_{u}^{\prime} = [v_1,\dots,v_{T},v_{1}^{\prime},\dots, v_{L}^{\prime}]$. This results in an attacked dataset $\mathcal{X}^{\prime} = \{\bm{x}_{u}^{\prime}|u\in\mathcal{U}\}$.
\item \textbf{White-box Attacks}: The attacker knows the sequential model $f_\theta$, as well as the training loss in Equation~\eqref{eq:training-loss}.

\item \textbf{Limited Insertion}: The amount of injected items $K$ is controlled to ensure that the recommendation system can maintain good performance after the attack.

\item \textbf{Attack without Delay}: The attacker can finish manipulating the training data before the recommender system starts training. Therefore, the sequential model parameters of $f_{\theta}$ will indeed be learned by fitting $\mathcal{X}^{\prime}$, \ie
$$\hat{\theta}^{\prime} = \arg\min_{\theta} \sum_{u\in\mathcal{U}} L(\bm{x}_{u}^{\prime};\theta),$$
where $\hat{\theta}^{\prime}$ denotes the learned model parameters based on $\mathcal{X}^{\prime}$, and $L(\cdot)$ is the same function used in Equation~\eqref{eq:training-loss}. 

\end{itemize}

\textit{Attack goal}: our attack goal is to find a $\mathcal{X}^{\prime}$ such that $f_{\hat{\theta}^{\prime}}$ can boost the recommendation of a target item $v^{*} \in \mathcal{V}$ as much as possible\footnote{Indeed, the attacker can also downgrade the recommendation of a target item. The goal can be achieved by promoting other items, so we only focus on promoting the target items in this work.}, compared to $f_{\hat{\theta}}$ trained with the original training data $\mathcal{X}$.

\section{Attack Plan Based on Influence Function}
In this section, we introduce the proposed attack algorithm. Firstly, we present the overall framework and then elaborate on the main components. Lastly, we provide a detailed algorithm.

\subsection{Overview}
\begin{figure}
    \centering
    \includegraphics[width=\linewidth]
    {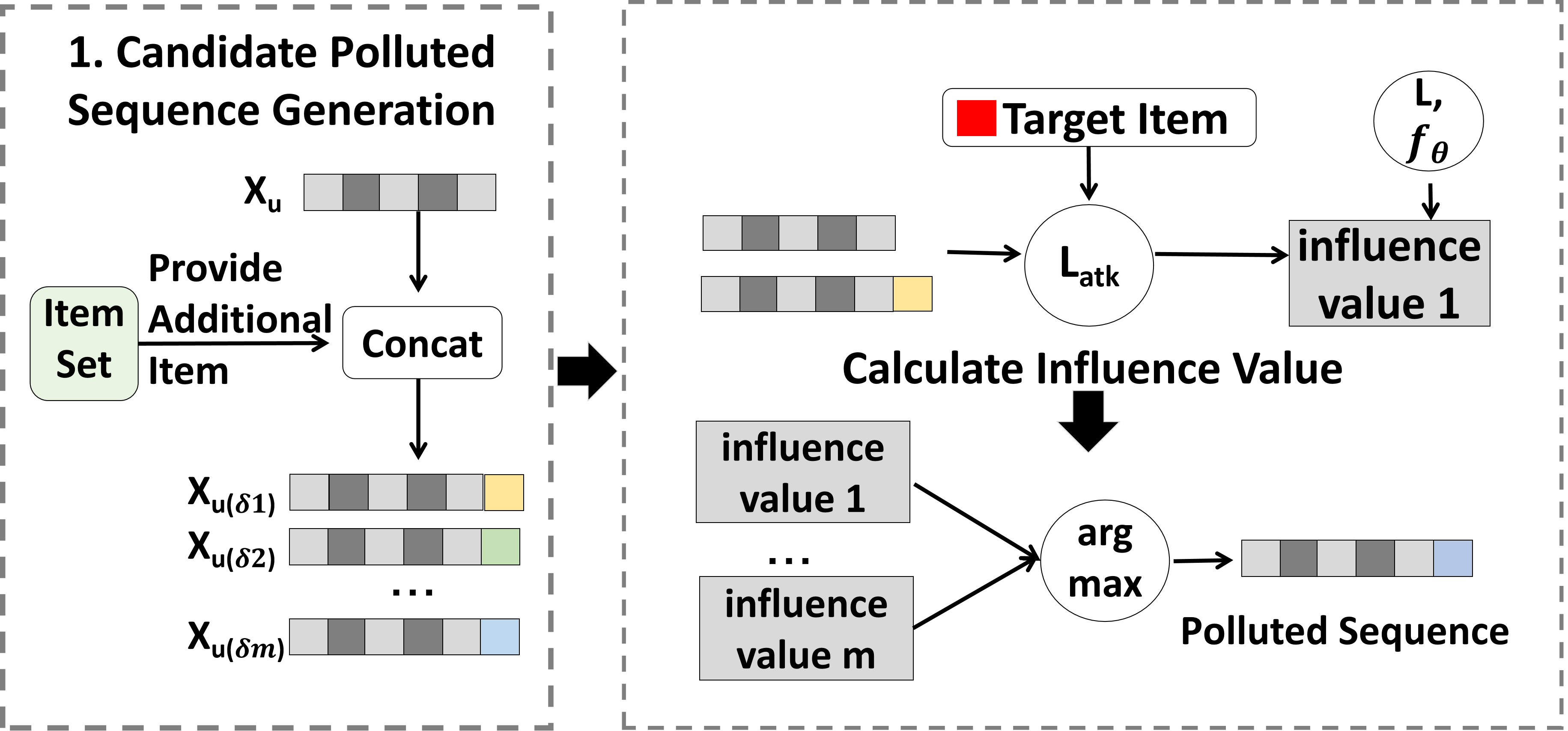}
    \caption{ The overall framework of the proposed INFAttack method, which mainly includes two parts: 1) candidate polluted sequence generation; and 2) polluted sequence screening.}
     \vspace{-2mm}
    \label{fig:attack-model}
    \vspace{-2mm}
\end{figure}

% To achieve our attack goal in the sequential recommendation, we propose an  \textit{\textbf{INF}luence function-based \textbf{Attack} } (INFAttack)  framework, which includes two main steps as shown in Figure~\ref{fig:attack-model}:
To achieve the defined attack goal, we propose an \textit{\textbf{INF}luence function-based \textbf{Attack}} (INFAttack) framework, which iteratively inserts \(K\) items. In each iteration, the most influential item at that time is injected using the following two main steps, as shown in Figure~\ref{fig:attack-model}:
\begin{itemize}[leftmargin=*]

   % \item Step 1: Candidate polluted sequence generation (the left box of Figure~\ref{fig:attack-model}). We first enumerate all possible combinations of injected items into one item, generating $m$ ($m=|\mathcal{V}|$) possible combinations. Then, for each original sequence $\bm{x}{u} \in \mathcal{X}$, we inject every possible combination into $\bm{x}{u}$, forming m contaminations sequence. 
   % Then, by filtering the pollution sequence, the optimal pollution sequence is obtained as the initial pollution sequence, and this step is repeated K times to obtain the final pollution sequence.
   \item Step 1: Candidate polluted sequence generation (the left box of Figure~\ref{fig:attack-model}). For each user's interaction sequence \(\bm{x}_{u} \in \mathcal{X}\), we enumerate all items and combine each possible item with \(\bm{x}_{u}\), forming \(m\) possible (intermediately) polluted sequences for \(\bm{x}_{u}\).

     % \item Step 1: Candidate polluted sequence generation (the left box of Figure~\ref{fig:attack-model}).  We first enumerate all possible combinations of the $K$ injected items, generating $m$ ($m=|\mathcal{V}|$) possible combinations. Then, for each original sequence $\bm{x}_{u} \in \mathcal{X}$, we inject each possible combination into $\bm{x}_{u}$, forming all possible polluted sequences for it. 

    % \item Step 2: Polluted sequence screening. The objective of this step is to select appropriate pollution sequences for users, thereby constructing the attacked dataset $\mathcal{X}^{\prime}$ and ensuring the efficacy of the attack. 
    % However, directly identifying the optimal combination of pollution sequences for all users proves arduous as a result of the vast selection space of $m*n$ ($n=|\mathcal{U}|$). Therefore, we adopt a greedy approach~\cite{DBLP:conf/kdd/ZugnerAG18} to execute this phase, which entails selecting the most influential pollution sequence for each user in terms of boosting the target item, instead of immediately determining the best composition of pollution sequences across all users. The resultant individual pollution sequences are then integrated into $\mathcal{X}^{\prime}$. The influential impact on the target item is gauged using the influence function~\cite{DBLP:conf/icml/KohL17}. The right box of Figure~\ref{fig:attack-model} shows the process.
    \item Step 2: Polluted sequence screening. The objective of this step is to select appropriate polluted sequences for users, ensuring the efficacy of the attack. Directly identifying the optimal combination of polluted sequences for all users is challenging due to the vast selection space of \(m^n\) (\(n = |\mathcal{U}|\)). Therefore, we adopt a greedy approach~\cite{DBLP:conf/kdd/ZugnerAG18}. This approach involves selecting the most influential polluted sequence for each user in terms of boosting the target item, rather than immediately determining the best composition of polluted sequences across all users.  The impact on the target item is gauged using the influence function~\cite{DBLP:conf/icml/KohL17}. The right box of Figure~\ref{fig:attack-model} illustrates this process.
\end{itemize}

The above two steps are iterated until \(K\) items have been injected for each user. The most influential polluted sequences selected in the final step construct the final attacked dataset \(\mathcal{X}^{\prime}\). Unlike gradient-based methods, which experience diminished attack efficacy when converting gradient-based attacks into real-world item sequences~\cite{DBLP:conf/recsys/YueHZM21, DBLP:conf/recsys/YueZKSW22}, INFAttack directly manipulates genuine items to generate polluted sequences, thereby ensuring optimal attack effectiveness for actual item sequences\footnote{At a local level.}. Next, we elaborate on the two main steps.

% In summary, our approach entails an exhaustive enumeration of all feasible polluted sequences, followed by a meticulous selection process using a greedy approach based on the influence function to prioritize the most influential ones for enhancing the recommendation of the target item, culminating in the creation of the final attacked dataset. Unlike gradient-based methods that experience diminished attack efficacy during the conversion of gradient-based attacks into real-world item sequences~\cite{DBLP:conf/recsys/YueHZM21, DBLP:conf/recsys/YueZKSW22}, INFAttack directly manipulates genuine items to generate polluted sequences, thereby ensuring optimal attack effectiveness for actual item sequences\footnote{At a local level.}. We next elaborate on the two main steps.

\subsection{Candidate Polluted Sequence Generation}

% As previously mentioned, we inject $K$ items to construct a polluted sequence. To generate all possible candidate polluted sequences, we first enumerated all possible combinations of $[v_{1}^{\prime},\dots,v_{K}^{\prime}]$ (\ie the $K$ injected items) from the item set $\mathcal{V}$, generating $m$ ($m=|\mathcal{V}|$) possible combinations. Then, for a original sequence $\bm{x}_{u} \in \mathcal{X}$, we inject each possible $[v_{1}^{\prime},\dots,v_{K}^{\prime}]$ into $\bm{x}_{u}$, getting the polluted sequence $\bm{x}_{u(\delta)}=[v_{1},\dots,v_{T},v_{1}^{\prime},\dots,v_{K}^{\prime}]$. Finally, for each $\bm{x}_{u}$, this process results in the candidate set of polluted sequences, denoted as $\{\bm{x}_{u(\delta_{p})}|p=1,\dots,m\}$ for user $u$, where $\bm{x}_{u(\delta_{p})}$ denotes the $p$-th possible polluted sequence. As shown in Figure~\ref{fig:method-Tvalue}, for each $\bm{x}_{u}$, there are $m$ possible candidate polluted sequences when $K=1$. By calculating the optimal sequence when k=1, repeat the process to calculate $K=2$. There are $m$ possible candidate polluted sequences when $K=2$.  

\begin{figure}[t]
    \centering
    \includegraphics[width=\linewidth]{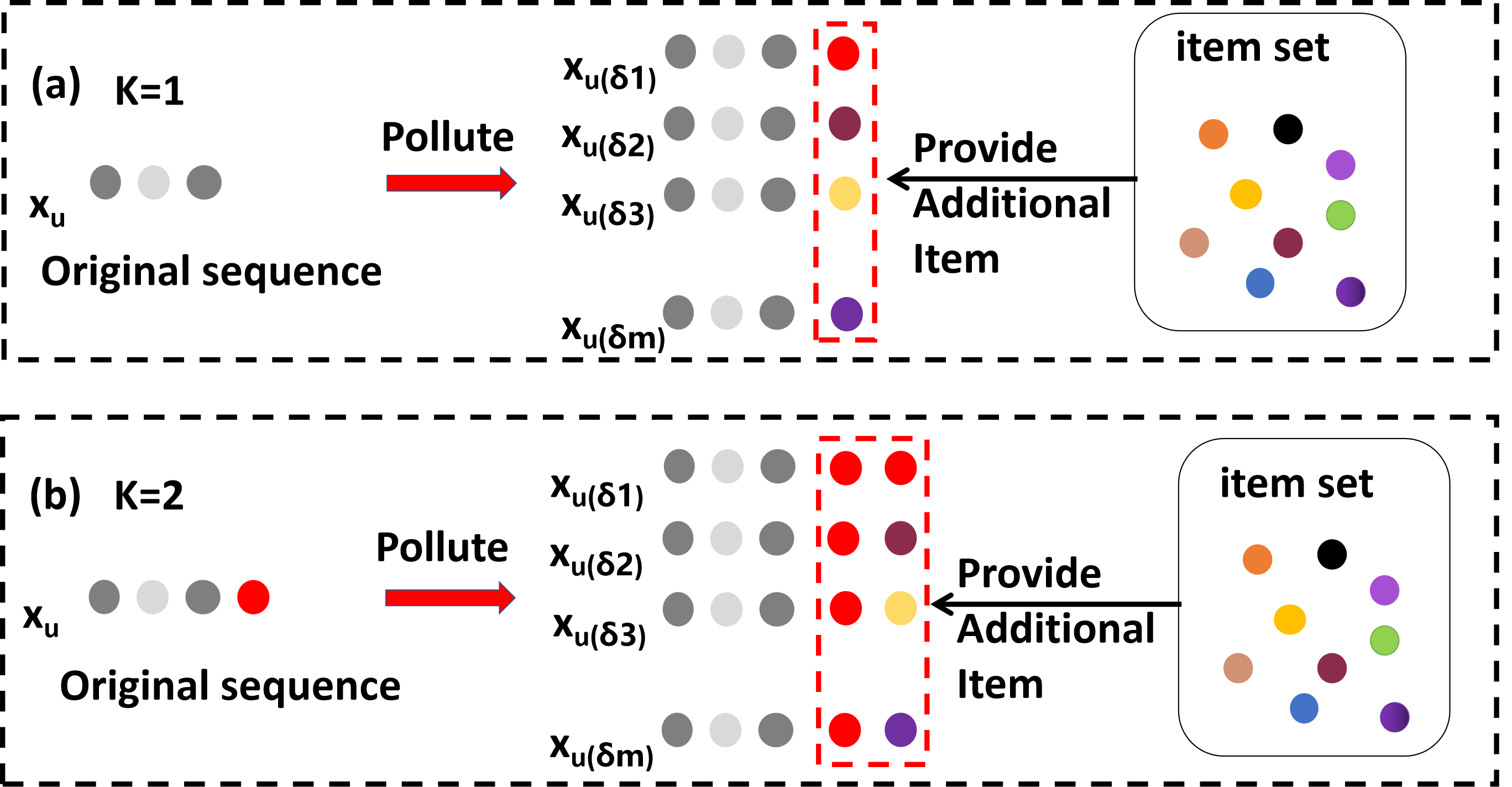}
    \caption{Illustration of generating candidate polluted sequences: (a) the case of injecting $1$ item, resulting in $m$ candidate polluted sequences; (b) the case of injecting $2$ items, resulting in $m$ candidate polluted sequences.}
    \vspace{-2mm}
    \label{fig:method-Tvalue}
     \vspace{-2mm}
\end{figure}

As previously mentioned, our goal is to inject \(K\) items to construct a polluted sequence. Finding an optimal combination of \(K\) suitable items \([v_{1}^{\prime}, \dots, v_{K}^{\prime}]\) at a time is challenging due to the large combination space. To address this, we choose to inject one item at a time. The candidate generation process identifies all possible polluted sequences with one injected item for all users. For each \(\bm{x}_u\), we inject the \(p\)-th item \(v \in \mathcal{V}\) into it, resulting in a polluted sequence \(\bm{x}_{u(\delta_{p})} = [\bm{x}_u, v]\). By injecting all possible items, we obtain a candidate set of polluted sequences for user \(u\), denoted as \(\{\bm{x}_{u(\delta_{p})} \mid p = 1, \dots, m\}\). As shown in Figure~\ref{fig:method-Tvalue}, for each \(\bm{x}_u\), there are \(m\) possible candidate polluted sequences. Notably, for the \(k+1\)-th injection, \(\bm{x}_u\) is the final result of the \(k\)-th injection, which will be elaborated on later.

\subsection{Polluted Sequence Screening}

% After generating all possible candidate polluted sequences for each $\bm{x}_{u}\in\mathcal{X}$, we need to screen out the polluted sequences that are most influential on enhancing the recommendation of the target item $v^{x}$. To avoid excessive computational overhead, we greedily screen out the most influential polluted sequence for one user each time, and then combine the polluted sequences screened out for all users to form the final attacked training data $\mathcal{X}^{\prime}$. We next concentrate on how to screen out the most influential polluted sequence for each user. 

After generating all possible candidate polluted sequences for each $\bm{x}_{u}\in\mathcal{X}$, we need to screen out the polluted sequences that are most influential on enhancing the recommendation of the target item $v^{x}$. To avoid excessive computational overhead, we greedily screen out the most influential polluted sequence for one user each time, and then combine the polluted sequences screened out for all users to form the final attacked sequence for this step's item injection. We next concentrate on how to screen out the most influential polluted sequence for each user. 

% Toward the goal, we first need to quantify the influence of each polluted sequence on enhancing the recommendation of target item $v^{*}$. To quantify the influence of a polluted sequence $\bm{x}_{u(\delta_{p})}$, the most straightforward way is to compare the recommendation results between the original model and the model obtained by replacing $\bm{x}_{u}$ with $\bm{x}_{u(\delta_{p})}$. However, directly retraining the model could be inefficient. We thus consider utilizing the influence function to estimate the influence with a similar idea but without explicitly retraining. We next first define the influence via the influence function and then present how to calculate it, followed by how to filter the most influential polluted sequences.
Toward the goal, we need to quantify the influence of each polluted sequence on enhancing the recommendation of target item $v^{*}$. To quantify the influence of a polluted sequence $\bm{x}_{u(\delta_{p})}$, the most straightforward way is to compare the recommendation results between the original model and the model obtained by replacing $\bm{x}_{u}$ with $\bm{x}_{u(\delta_{p})}$. However, directly retraining the model could be inefficient. We thus consider utilizing the influence function to estimate the influence with a similar idea but without explicitly retraining. We next first define the influence via the influence function and then present how to calculate it, followed by how to select the most influential polluted sequences.

\vspace{+5pt}
\subsubsection{Influence of Polluted Sequence on the Target Item.} We first present the basic definition of the utilized influence, and then introduce its detailed expression.

\noindent \textbf{Basic definition.}
For the original model and the model gotten by replacing $\bm{x}_{u}$ with $\bm{x}_{u(\delta_{p})}$, we could represent them with a uniform formulation by adding the loss difference of fitting $\bm{x}_{u}$ and $\bm{x}_{u(\delta_{p})}$ into the original training loss with a weight $\epsilon$ in Equation~\eqref{eq:training-loss} as follows:

\begin{equation}\small
\begin{split}
\label{params}
    \begin{aligned}
        \hat{\theta}_{\epsilon,x_u,x_{u(\delta p)}}=\mathop{\arg\min}\limits_{\theta} \frac{1}{n}\sum_{u^{\prime}\in\mathcal{U}} L(\bm{x}_{u^{\prime}};\theta) \\ +  \epsilon (L(\bm{x}_{u(\delta p)};\theta)- L(\bm{x}_u;\theta)).
    \end{aligned}
    \end{split}
\end{equation}
The first term corresponds to the training loss on $\mathcal{X}$, identical to Equation~\eqref{eq:training-loss}. The terms $L(\bm{x}_{u(\delta p)},\theta)$ and $L(\bm{x}_u,\theta)$ represent the fitting losses for $\bm{x}_{u(\delta p)}$ and $\bm{x}_{u}$ with the model parameterized by $\theta$, respectively. It is easily verified that: 1) $\hat{\theta}_{\epsilon=0,x_u,x_{u(\delta p)}}$ corresponds to the original model, \ie $\hat{\theta} = \hat{\theta}_{\epsilon=0,x_u,x_{u(\delta p)}}$; and 2) $\hat{\theta}_{\epsilon=\frac{1}{n},x_u,x_{u(\delta p)}}$ corresponds to the model gotten by  replacing $\bm{x}_{u}$ with $\bm{x}_{u(\delta_{p})}$. 

To quantify the influence of $\bm{x}_{u}$ on enhancing the recommendation of target item $v^{*}$, we only need to compare the difference between $\hat{\theta}_{\epsilon=\frac{1}{n},x_u,x_{u(\delta p)}}$ and  $\hat{\theta}_{\epsilon=0,x_u,x_{u(\delta p)}}$ on recommending the target item $v^{*}$. We formalize the difference with a cross-entropy loss as follows:

\begin{equation}\small
\begin{split}
     R(\bm{x}_{u(\delta)} \leftarrow \bm{x}_{u}) = L_{atk}(\hat{\theta}_{\epsilon=\frac{1}{n},x_u,x_{u(\delta p)}};v^{*})    \\ - L_{atk}(\hat{\theta}_{\epsilon=0,x_u,x_{u(\delta p)}};v^{*}),
\end{split}
\end{equation}
where we leverage $R(\bm{x}_{u(\delta)} \leftarrow \bm{x}_{u})$ to denote the difference, and $L_{atk}(\hat{\theta}_{\epsilon=\frac{1}{n},x_u,x_{u(\delta p)}};v^{*})$ denotes how likely the model with $\hat{\theta}_{\epsilon=\frac{1}{n},x_u,x_{u(\delta p)}}$ would recommend the target item $v^{*}$, which is computed according to the cross-entropy loss, formally,
\begin{equation} \small 
\label{L_{atk}}
\begin{aligned}
L_{atk}(\hat{\theta}_{\epsilon=\frac{1}{n},x_u,x_{u(\delta p)}};v^{*}) = \sum_{u} L_{ce} (f_{\hat{\theta}_{\epsilon=\frac{1}{n},x_u,x_{u(\delta p)}}}(\bm{x}_{u});v^{*}),
\end{aligned}
\end{equation}
where $L_{ce}(\cdot;v^{*})$ denotes the cross-entropy loss regarding the target item $v^{*}$, with lower value denoting a higher probability to recommending $v^{*}$. Then, according to Taylor's series, we have:

\begin{equation} \small
    \begin{split}
        R(\bm{x}_{u(\delta)} \leftarrow \bm{x}_{u}) &\approx \frac{\partial L_{atk}(\hat{\theta}_{\epsilon,x_u,x_{u(\delta p)}};v^{*})  }{\partial \epsilon} \bigg|_{\epsilon=0} \times \frac{1}{n}\\
        & \propto \frac{\partial L_{atk}(\hat{\theta}_{\epsilon,x_u,x_{u(\delta p)}};v^{*})  }{\partial \epsilon} \bigg|_{\epsilon=0}
    \end{split}.
\end{equation}
This equation means we could define the influence of $\bm{x}_{u(\delta)}$ on promoting the target item $v^{*}$ as:
\begin{equation} \small
\label{I-atk(noparam)}
    \begin{aligned}
        I_{\epsilon,atk}(x_u,x_{u(\delta p)}) \overset{def}{=} \frac{\partial L_{atk} (\hat{\theta}_{\epsilon,x_u,x_{u(\delta p)}};v^{*})}{\partial \epsilon} \bigg\rvert_{\epsilon=0}
    \end{aligned},
\end{equation}
for which a higher value means a higher influence on enhancing the recommendation of $v^{*}$. We next introduce how to further derive it to make it calculable.

\vspace{+5pt}
\noindent \textbf{Detailed expression.} We have defined the influence of a polluted sequence $\bm{x}_{u(\delta)}$ on enhancing the recommendation of $v^{*}$ based on $\hat{\theta}_{\epsilon,x_u,x_{u(\delta p)}}$ in Equation~\eqref{params}. We next further derive it such that it could be easily calculated according to the model parameters $\hat{\theta}$ trained with the original dataset. According to the Chain rule, we have:
\begin{small}
\begin{equation}
\label{eq:chin-rule}
    \begin{aligned}
        I_{\epsilon,atk}(x_u,x_{u(\delta p)}) & {=} \frac{\partial L_{atk} (\hat{\theta}_{\epsilon,x_u,x_{u(\delta p)}};v^{*})}{\partial \epsilon} \bigg\rvert_{\epsilon=0}\\
        & = \nabla_{\theta} L_{atk} (\hat{\theta}; v^{*}) ^ \top  \frac{\partial \hat{\theta}_{\epsilon,x_u,x_{u(\delta p)}}}{\partial \epsilon} \bigg\rvert_{\epsilon=0}
    \end{aligned},
\end{equation}
\end{small}
where the second term $I_{\epsilon,p}=\frac{\partial \hat{\theta}_{\epsilon,x_u,x_{u(\delta p)}}}{\partial \epsilon} \bigg\rvert_{\epsilon=0}$ is the influence of $\bm{x}_{u}$ on the model parameters. According to the classical results~\cite{DBLP:conf/www/FangG020}, $I_{\epsilon,p}$ can be expressed as follows:

% \begin{equation}\small
% \begin{split}
% \label{I-theta}
% \begin{aligned}
%     I_{\epsilon,p} & = \frac{\partial \hat{\theta}_{\epsilon,x_u,x_{u(\delta p)}}}{\partial \epsilon} \bigg\rvert_{\epsilon=0}
%      = -(H_{\hat{\theta}}+\lambda{I})^{-1} \\ &\quad  *\nabla_{\theta} \big(L(x_{u(\delta p)};\hat{\theta})- L(x_u;\hat{\theta})\big),
%     % \\ &\approx -(H_{\theta}+\lambda{I})^{-1}\nabla_{x_u} \nabla_{\theta} L(x_u,\theta))(x_{u(\delta p)})-(x_u).
% \end{aligned}
% \end{split}
% \end{equation}

\begin{equation}\small
\label{I-theta}
\begin{aligned}
    I_{\epsilon,p} = \frac{\partial \hat{\theta}_{\epsilon,x_u,x_{u(\delta p)}}}{\partial \epsilon} \bigg\rvert_{\epsilon=0}
     = -(H_{\hat{\theta}}+\lambda{I})^{-1}  *\nabla_{\theta} \left(L(x_{u(\delta p)};\hat{\theta})- L(x_u;\hat{\theta})\right),
    % \\ &\approx -(H_{\theta}+\lambda{I})^{-1}\nabla_{x_u} \nabla_{\theta} L(x_u,\theta))(x_{u(\delta p)})-(x_u).
\end{aligned}
\end{equation}

where $H_{\hat{\theta}} = \frac{1}{n}\sum_{u^{\prime}\in\mathcal{U}} \nabla_{\theta}^2 L(x_{u^{\prime}};\hat{\theta})$ is the Hessian matrix, $L(\cdot)$ denotes the loss defined in Equation~\eqref{eq:training-loss}, $I$ is a identity matrix, and $\lambda$ is a weighting coefficient.  Note that for the original form of the influence function in~\cite{DBLP:conf/icml/KohL17}, there is not a term of $\lambda I$, \ie $\lambda=0$, with assuming that $H_{\hat{\theta}}$  is positive definite. However, the assumption is not usually satisfied for sequential models, we thus follow previous work~\cite{DBLP:journals/corr/abs-2209-05364} to add the term of $\lambda I$, making $H_{\hat{\theta}}+\lambda I$ invertible.

\textit{{Final expression.}} According to Equation \eqref{I-theta} and Equation~\eqref{eq:chin-rule}, the influence $I_{\epsilon,atk}(x_u, x_{u(\delta p) )})$ could be fully computed with only $\hat{\theta}$ and the defined training loss function $L(\cdot)$ and $L_{atk}(\cdot)$ as follows:

\begin{equation}\small
\begin{split}
\label{I-atk(param)}
    \begin{aligned}
        I_{\epsilon,atk}(x_u,x_{u(\delta p)}) & = -\nabla_{\theta}L_{atk}(\hat{\theta};v^{*})^T (H_{\hat{\theta}}+\lambda{I})^{-1} \\ &\quad  * \nabla_{\theta}\big(L(x_{u(\delta p)};\hat{\theta})- L(x_u;\hat{\theta})\big)     
    \end{aligned}.
\end{split}
\end{equation}

 \vspace{+5pt}
\subsubsection{Calculate the Influence.}   
Estimating $I_{\epsilon,atk}(x_u,x_{u(\delta p)})$ involves forming and inverting the $H_{\hat{\theta}}+\lambda I$, which could  be computation-costly. 
%Hessian matrix $H_{\hat{\theta}}$ regarding the training loss,
To efficently computing $H_{\theta}$ explicitly, we approximate $H_{\theta}^{-1}$ using the  stochastic estimation proposed in   ~\cite{DBLP:conf/icml/KohL17}. For clarity, let $\Tilde{H} = H_{\hat{\theta}}+\lambda I$, and let $\Tilde{H}_{j}^{-1} = \sum_{i=1}^{j}(I-\Tilde{H}_{j})^{i}$, the first $j$-th terms of the Taylor expansion of $H^{-1}$. And we can re-write $\tilde{H}_{j}^{-1}$ in a recursive form as: $\tilde{H}_{j}^{-1}= I+ (I-\tilde{H}) \tilde{H}_{j-1}^{-1}$. Then it is easily verified that   $\Tilde{H}_{j}^{-1} \rightarrow \Tilde{H}^{-1}$. 

Particularly, at each iteration $j$, we can sample a user $u$, and take $\nabla_{\theta}^{2} L(x_{u};\hat{\theta})$ as an unbiased estimation of the full $\tilde{H}$. After running $J$ interactions, we could approach $\tilde{H}$ with  $\Tilde{H}_{J}^{-1}$. We could run the process multiple times, and take the average of $\Tilde{H}_{J}^{-1}$ as a better estimation for $\tilde{H}$, still denoted as $\tilde{H}_{J}^{-1}$. Finally, we could compute $I_{\epsilon,atk}(x_u,x_{u(\delta p)})$ as follows:

\begin{equation}\small
\begin{split}
\label{I-atk-compute}
    I_{\epsilon,atk}(x_u,x_{u(\delta p)}) & = -\nabla_{\theta}L_{atk}(\hat{\theta};v^{*})^T \tilde{H}_{J}^{-1} \\ &\quad *\nabla_{\theta}\big(L(x_{u(\delta p)};\hat{\theta}) - L(x_u;\hat{\theta})\big).
\end{split}
\end{equation}

Therefore, to estimate $I_{\epsilon,atk}(x_u,x_{u(\delta p)})$, the stochastic estimation method avoids the explicit calculation of the inverse of the Hessian matrix, which accelerates the calculation of the influence value.

\subsubsection{Filtering}
% Finally, for each candidate polluted sequence $\bm{x}_{u(\delta_{p})}$ of the user $u$, we could compute its influence as $I_{\epsilon,atk}(x_u,x_{u(\delta_{p})})$. 
% Then, we select the one with the largest absolute value of influence as the final pollution sequence, as shown below:
% % Then, we select the one with the highest influence as the final polluted sequence as follows:
% $$x_{u}^{\prime} = \bm{x}_{u(\delta_{p^{\prime}})},\quad p^\prime = argmax_{p\in\{1,\dots,m\}} |I_{\epsilon,atk}(x_u, x_{u(\delta_{p})})|.$$
% Lastly, the final attack training data $\mathcal{X}^{\prime}$ is constructed by combining all $\bm{x}_{u}^{\prime}$, \ie  $\mathcal{X}^{\prime}=\{\bm{x}_{u}^{\prime}|u\in \mathcal{U}\}$.

Finally, after obtaining the influence \(I_{\epsilon,atk}(\bm{x}_u, \bm{x}_{u(\delta_{p})})\) for each candidate polluted sequence \(\bm{x}_{u(\delta_{p})}\), we select the one with the largest absolute value of influence as the final pollution sequence for the current step of injection, as shown below:
$$\quad p^\prime = \arg\max_{p \in \{1, \dots, m\}} |I_{\epsilon,atk}(\bm{x}_u, \bm{x}_{u(\delta_{p})})|,$$
where \(p^\prime\) denotes the index of the most influential polluted sequence. If the number of injected items is less than \(K\), we treat \(\bm{x}_{u(\delta_{p^\prime})}\) as the new \(\bm{x}_u\). Otherwise, if this is the last step for injecting items, we treat \(\bm{x}_{u(\delta_{p^\prime})}\) as the final polluted sequence \(\bm{x}_{u}^{\prime}\) for \(u\), and construct the final attack training data \(\mathcal{X}^{\prime}\) by combining all \(\bm{x}_{u}^{\prime}\), i.e., \(\mathcal{X}^{\prime} = \{\bm{x}_{u}^{\prime} \mid u \in \mathcal{U}\}\).

\subsection{Constructing the Attack Process}
% Our proposed algorithm can pollute the recommender system by injecting the most influential items to perturb the sequence of users. Ideally, the contamination sequence leads to increased exposure to the target item. To make the attack algorithm clearer, the detailed steps of the attack are described in Algorithm(\ref{alg:INF-atk}).

The INFAttck method aims to manipulate recommender systems by contaminating user interaction sequences to increase exposure to a specific item. To provide a clear understanding of the algorithm, we present Algorithm (\ref{alg:INF-atk}) with detailed attack steps. Firstly, we train the sequential recommender model $f_{\theta}$ on original training data $\mathbf{X}$ to obtain the model parameters $\theta$ (line 2), which are required for the influence computation. Then, we generate the most influential polluted sequence for each user to construct the attacked data $\mathcal{X}^{\prime}$ (lines 3-13). For each user, we cycle $K $times to select the most suitable injection items $ {x_ {u ( delta_ {p})} | p=1,  dots, m } $(lines 4-6), calculate the defined impact of each candidate pollution sequence on the target item (lines 8-10) and ultimately generate the final pollution sequence of user $u $as the sequence with the greatest impact among all candidate pollution sequences (lines 11-12)

% \begin{algorithm}[t]
%     \caption{INFAttack}
%     \LinesNumbered
%     \label{alg:INF-atk}
%     \KwIn{Original training sequences $\mathcal{X}$, target item $v^{*}$, recommender model $f_{\theta}$, the number of injected items $K$, and the item set $\mathcal{V}$}
%     \KwOut{A set of polluted sequences $\mathcal{X}^{\prime}$}
%     Initialize $\mathcal{X}^{\prime}=\varnothing$\;
%     Train $f_{\theta}$ with $\mathcal{X}$ to obtain $\hat{\theta}$ according to Equation~\eqref{eq:training-loss}\;
%     \For{each $x_{u} \in \mathcal{X}$}{
%         \While{length of $x_{u(\delta_{p})} $ - length of $x_u < K$}{
%             // Step 1: Generate all candidate polluted sequences\;
%             Enumerate $m$ possible combinations of $v_{i}^{\prime}$ ($i \leq K$) from $\mathcal{V}$\;
%             Inject each $[v_{i}^{\prime}]$ into $\bm{x}_{u}$, obtaining $\{x_{u(\delta_{p})}|p=1,\dots,m\}$\;
%             // Step 2: Filter pollution sequences\;
%             Compute $I_{\epsilon,atk}(x_u,x_{u(\delta_p)})$ with $\hat{\theta}$ according to Equation~\eqref{I-atk-compute}\;
%             \If{max($I_{\epsilon,atk}(x_u,x_{u(\delta_p)})) < 0$}
%             {
%               \STATE break;
%             }              
%             \Else
%                { Find $p^{\prime} = \text{argmax}_{p} \, |I_{\epsilon,atk}(x_u,x_{u(\delta_p)})|$\;
%                 $x_{u} = x_{u(\delta_{p})}$\;
%                 Append $x_{u(\delta_{p^{\prime}})}$ into $\mathcal{X}^{\prime}$\;
%                }
            
%         }
%     }
%     % \KwResult{$\mathcal{X}^{\prime}$}
%   \Return $\mathcal{X}^{\prime}$
% \end{algorithm}

\begin{algorithm}[t]
    \caption{INFAttack}
    \LinesNumbered
    \label{alg:INF-atk}
    \KwIn{Original training sequences $\mathcal{X}$, target item $v^{*}$, recommender model $f_{\theta}$, number of injected items $K$, and item set $\mathcal{V}$}
    \KwOut{A set of polluted sequences $\mathcal{X}^{\prime}$}
    
    Initialize $\mathcal{X}^{\prime} = \varnothing$\;
    Train $f_{\theta}$ with $\mathcal{X}$ to obtain $\hat{\theta}$ according to Equation~\eqref{eq:training-loss}\;
    
    \For{each $x_{u} \in \mathcal{X}$}{
        \While{length of $x_{u(\delta_{p})} - \text{length of } x_u < K$}{
            \tcc{Step 1: Generate all candidate polluted sequences}
            Enumerate $m$ possible combinations of $v_{i}^{\prime}$ ($i \leq K$) from $\mathcal{V}$\;
            Inject each $[v_{i}^{\prime}]$ into $\bm{x}_{u}$, obtaining $\{x_{u(\delta_{p})}|p=1,\dots,m\}$\;
            
            \tcc{Step 2: Filter pollution sequences}
            Compute $I_{\epsilon,atk}(x_u,x_{u(\delta_p)})$ with $\hat{\theta}$ according to Equation~\eqref{I-atk-compute}\;
            
            \If{max($I_{\epsilon,atk}(x_u,x_{u(\delta_p)}) < 0$)}{
                \textbf{break}\;
            }
            \Else{
                Find $p^{\prime} = \text{argmax}_{p} \, |I_{\epsilon,atk}(x_u,x_{u(\delta_p)})|$\;
                $x_{u} = x_{u(\delta_{p})}$\;
                Append $x_{u(\delta_{p^{\prime}})}$ into $\mathcal{X}^{\prime}$\;
            }
        }
    }
    \KwRet $\mathcal{X}^{\prime}$\;
\end{algorithm}

% \begin{algorithm}[t]
%     \caption{INFAttack}
%     \LinesNumbered
%     \label{alg:INF-atk}
%     \KwIn{Original training sequences $\mathcal{X}$, target item $v^{*}$, recommender model $f_{\theta}$, the number of injected items $K$, and the item set $\mathcal{V}$}
%     \KwOut{A set of polluted sequences $\mathcal{X}^{\prime}$}
%     Initialize $\mathcal{X}^{\prime}=\varnothing$\;
%     Train $f_{\theta}$ with $\mathcal{X}$ to obtain $\hat{\theta}$ according to Equation~\eqref{eq:training-loss}\;
%     \For{each $x_{u} \in \mathcal{X}$}{
%     \While{length of $x_{u(\delta_{p})} $-length of $x_u < K$}{
%    // Step 1: Generate all candidate polluted sequence\;
%     Enumerate $m$ possible combinations of $v_{i}^{\prime}(i<=K)$ from $\mathcal{V}$\;
%     Inject each $[v_{i}^{\prime}]$ into $\bm{x}_{u}$, getting $\{x_{u(\delta_{p})}|p=1,\dots,m\}$\;
%     // Step 2:  Filter pollution sequences\;
   
%     Compute  $I_{\epsilon,atk}(x_u,x_{u(\delta_p)})$ with $\hat{\theta}$ according to Equation~\eqref{I-atk-compute}\; 
%     \IF{max($I_{\epsilon,atk}(x_u,x_{u(\delta_p)})$)<0}
% 		\STATE  break;
%     \ELSE
% 		Find $p^{\prime} = argmax_{p} \, I_{\epsilon,atk}(x_u,x_{u(\delta_p)})$\;
%         $x_{u}=x_{u(\delta_{p})}$;
%         Append $x_{u(\delta_{p^{\prime}})}$ into $\mathcal{X}^{\prime}$\;
%     \ENDIF
    
%     }   
%     } 
%     Return $\mathcal{X}^{\prime}$
 
% \end{algorithm}  

\section{EXPERIMENTS}
We conducted extensive experiments to answer the following research questions:
\textbf{RQ1}: How effective is our INFAttack in achieving profile pollution attack compared to existing SOTA methods?  \textbf{RQ2}: Will our attack method significantly affect the recommendation performance of the recommender system? \textbf{RQ3}: How important is the influence function for the effectiveness of our proposal? \textbf{RQ4}: How do hyperparameters (such as K) affect the attack and recommendation performance of the proposed methods?

% \noindent\textbf{RQ1}: How effective is our INFAttack in achieving profile pollution attack compared to existing SOTA methods?

% \noindent\textbf{RQ2}: Will our attack method significantly affect the recommendation performance of the recommender system?

% \noindent\textbf{RQ3}: How important is the influence function for the effectiveness of our proposal?

% \noindent\textbf{RQ4}: How do hyperparameters (such as K) affect the attack and recommendation performance of the proposed methods?

% \noindent\textbf{RQ5}: How efficient is each mode in implementing pollution attacks in terms of operational effectiveness?

% \noindent\textbf{RQ1}: Can We Attack Recommenders using Influence-based Profile Pollution? 

% \noindent\textbf{RQ2}: How does impact-based profile pollution affect the performance of the recommender system?

% \noindent\textbf{RQ3}: What role does our proposed influence play in the attack?

% \noindent\textbf{RQ4}: What impact does the change of $K$ value have on the attack effect and recommendation effect?

\subsection{Experimental Settings}
\subsubsection{Datasets}
We evaluate our proposed method using five commonly used benchmark datasets: Movielens (both ML-1M and ML-20M)~\cite{DBLP:journals/tiis/HarperK16}, Steam ~\cite{DBLP:conf/sigir/McAuleyTSH15}, Beauty~\cite{DBLP:conf/emnlp/NiLM19} and LastFM~\cite{2008Music}. Following the preprocessing method in BERT4Rec~\cite{DBLP:conf/cikm/SunLWPLOJ19}, we handle the rating data as implicit feedback. We follow the data processing requirements of SASRec and BERT4Rec, using the last two items in each sequence for validation and testing, and the rest for training. 
We summarize the statistics of datasets in Table~\ref{tab:DataSets}.

\begin{table}[ht]
  \caption{The statistics of datasets. 'Avg. Len.' denotes the "average length" of user interaction sequences. }
  \vspace{-5pt}
  \label{tab:DataSets}
  \begin{tabular}{ccccl}
    \toprule
    Dataset & \#Users & \#Items & Avg. Len. & Sparsity\\
    \midrule
    ML-1M  & 6,040   & 3,416   & 165 & 95.2\% \\
    Beauty & 22,363  & 12,101  &  9  & 99.9\% \\
    Steam  & 334,289 & 12,012  & 13  & 99.9\% \\
    LastFM &998      &57638    &9739 &83.1\% \\
    ML-20M &138,493   &18,345   & 13  &99.2\%\\
    \bottomrule
  \end{tabular}
  \vspace{-2mm}
\end{table}

\subsubsection{Backbone Sequential Recommender}
To evaluate our proposed attack method, we conduct experiments on four different sequential recommenders, including

\begin{itemize}[leftmargin=*]
    \item[-] \textbf{NARM~\cite{DBLP:conf/cikm/LiRCRLM17}:} This model uses two GRU encoders to capture both local and global sequential information.

    \item[-]\textbf{SASRec~\cite{DBLP:conf/icdm/KangM18}:} This is a sequential recommendation model based on one-way self attention, which uses previous interactions to predict future interactions during training.

    \item[-]\textbf{BERT4Rec~\cite{DBLP:conf/cikm/SunLWPLOJ19}:} This is a self-attention network-based model with a similar architecture to SASRec. However, it uses bidirectional self-attention.
    \item[-]\textbf{Locker~\cite{DBLP:conf/cikm/HeZ0WKM21}:} This is similar to BERT4Rec, but in our implementation, we use convolutional layers to model local dynamics.
\end{itemize}

\subsubsection{Evaluation Protocols}
We conducted top-N recommendations to evaluate both the attack and recommendation performance of all compared methods but with different metrics:

\begin{itemize}[leftmargin=*]
\item \textit{Attack evaluation.} To evaluate the attack effectiveness, we used two widely used metrics in the literature that quantify the effectiveness of the method in promoting the target item: Recall rate (R@N) and Normalized Discounted Cumulative Gain (NDCG@N) for the target item. R@N represents the proportion of users who received the target item in the top-N recommendation list. NDCG@N is an evaluation of the ranking of the target item in the recommendation list. A higher achieved R@K and NDCG@K indicate better effectiveness of the method in maximizing the promotion attack target.

\item \textit{Recommendation evaluation.} We used hit rate (HR@K) to measure whether the attack method will destroy the recommendation validation of the underlying RS. To do this, we used the hit rate (HR@K)~\cite{DBLP:journals/air/RezaimehrD21} to evaluate the accuracy of the RS with pollution attack. The higher this metric, the better the performance the model can achieve.
\end{itemize}

\subsubsection{Baselines}
As the proposed INFAttack method is a white-box profile pollution attack algorithm on sequential recommendation, we compared it with the following baselines that are also white-box profile pollution attack algorithms designed for sequential recommendation:
\begin{itemize}[leftmargin=*]
    \item[-] \textbf{RandomAlter}~\cite{DBLP:conf/wsdm/TangW18}: This method constructs pollution attacks by injecting randomly sampled items and target items into user interaction sequences.   
    % alternating interactions with random items and target item.  
    \item[-] \textbf{SimAlter}~\cite{DBLP:conf/recsys/YueHZM21}: This is a user profile pollution attack method in sequential recommendation that constructs attacks by injecting target items and items similar to the target items. Item similarity is calculated based on item embeddings.

\item[-] \textbf{Replace}~\cite{DBLP:conf/recsys/YueZKSW22}: 
% This is a SOTA gradient-based user profile pollution attack method, which takes the gradient of the attack loss (similar to $L_{atk}$ in Equation~\eqref{xxx}) to guide the choices of injected items.
 This is a state-of-the-art gradient-based user profile pollution attack method that utilizes the gradient of the attack loss (similar to $L_{atk}$ in Equation~\eqref{L_{atk}}) to guide the selection of injected items.

\end{itemize}

\begin{table*}[]
\caption{Comparisons of attack performance between different methods. The "Clean" column represents the performance of recommending target items without any attacks. The "SimAlter", "Replace" and "INFAttack" rows represent the performances under attacks using the corresponding methods.} 
\vspace{-10pt}

  \label{tab:attack-performance}
    \resizebox{0.9\linewidth}{4cm}{
\begin{tabular}{c|ccccccccccc}
\hline
\multirow{2}{*}{Rec.} & Dataset & \multicolumn{2}{c}{ML-1M} & \multicolumn{2}{c}{Steam} & \multicolumn{2}{c}{Beauty} & \multicolumn{2}{c}{LastFM} & \multicolumn{2}{c}{ML-20M} \\ \cline{3-12} 
 & Attck & N@10 & R@10 & N@10 & R@10 & N@10 & R@10 & N@10 & R@10 & N@10 & R@10 \\ \hline
\multirow{5}{*}{NARM} & Clean & 0.064 & 0.140 & 0.070 & 0.122 & 0.066 & 0.126 & 0.099 & 0.147 & 0.093 & 0.157 \\
 & RandomAlter & 0.069 & 0.155 & 0.092 & 0.128 & 0.072 & 0.138 & 0.100 & 0.146 & 0.095 & 0.161 \\
 & SimAlter & 0.073 & 0.165 & 0.091 & 0.148 & 0.082 & 0.158 & 0.100 & 0.148 & 0.098 & 0.165 \\
 & Replace & 0.215 & 0.316 & 0.141 & 0.253 & 0.214 & 0.320 & 0.262 & 0.407 & 0.140 & 0.223 \\
 & INFAttack & \textbf{0.274} & \textbf{0.467} & \textbf{0.272} & \textbf{0.352} & \textbf{0.256} & \textbf{0.372} & \textbf{0.273} & \textbf{0.421} & \textbf{0.168} & \textbf{0.256} \\ \hline
\multirow{5}{*}{SASRec} & Clean & 0.065 & 0.135 & 0.073 & 0.129 & 0.069 & 0.103 & 0.152 & 0.164 & 0.135 & 0.173 \\
 & RandomAlter & 0.077 & 0.148 & 0.102 & 0.168 & 0.112 & 0.168 & 0.115 & 0.168 & 0.110 & 0.182 \\
 & SimAlter & 0.089 & 0.172 & 0.160 & 0.281 & 0.176 & 0.292 & 0.145 & 0.231 & 0.124 & 0.211 \\
 & Replace & 0.216 & 0.353 & 0.178 & 0.292 & 0.187 & 0.325 & 0.191 & 0.296 & 0.139 & 0.226 \\
 & INFAttack & \textbf{0.302} & \textbf{0.487} & \textbf{0.217} & \textbf{0.355} & \textbf{0.320} & \textbf{0.432} & \textbf{0.213} & \textbf{0.315} & \textbf{0.146} & \textbf{0.320} \\ \hline
\multirow{5}{*}{BERT4Rec} & Clean & 0.057 & 0.126 & 0.068 & 0.129 & 0.064 & 0.143 & 0.086 & 0.155 & 0.106 & 0.157 \\
 & RandomAlter & 0.077 & 0.156 & 0.072 & 0.136 & 0.069 & 0.163 & 0.087 & 0.155 & 0.106 & 0.156 \\
 & SimAlter & 0.072 & 0.149 & 0.085 & 0.150 & 0.074 & 0.170 & 0.088 & 0.156 & 0.106 & 0.158 \\
 & Replace & 0.114 & 0.183 & 0.134 & 0.234 & 0.145 & 0.298 & 0.114 & 0.184 & 0.119 & 0.181 \\
 & INFAttack & \textbf{0.240} & \textbf{0.364} & \textbf{0.254} & \textbf{0.278} & \textbf{0.182} & \textbf{0.345} & \textbf{0.125} & \textbf{0.205} & \textbf{0.125} & \textbf{0.200} \\ \hline
\multirow{5}{*}{Locker} & Clean & 0.073 & 0.147 & 0.097 & 0.159 & 0.065 & 0.130 & 0.104 & 0.157 & 0.084 & 0.150 \\
 & RandomAlter & 0.075 & 0.152 & 0.099 & 0.162 & 0.064 & 0.126 & 0.106 & 0.162 & 0.086 & 0.152 \\
 & SimAlter & 0.084 & 0.167 & 0.135 & 0.230 & 0.133 & 0.242 & 0.109 & 0.166 & 0.09 & 0.159 \\
 & Replace & 0.234 & 0.356 & 0.131 & 0.225 & 0.134 & 0.25 & 0.126 & 0.200 & 0.107 & 0.180 \\
 & INFAttack & \textbf{0.258} & \textbf{0.382} & \textbf{0.138} & \textbf{0.245} & \textbf{0.138} & \textbf{0.258} & \textbf{0.129} & \textbf{0.200} & \textbf{0.115} & \textbf{0.211} \\ \hline
\end{tabular}
}
\end{table*}

\subsubsection{Implementation Details}
To ensure a fair comparison, we followed the authors' implementations when implementing backbone models and baselines. All models were optimized using the Adam optimizer with a default learning rate of 0.001, weight decay of 0.01, and batch size of 64. We set the maximum sequence lengths for ML-1M, Steam, and Beauty datasets as 200, 50, and 50, respectively, following the settings of SASRec and BERT4Rec. Regarding the number of injected items (\ie $K$) when performing the attack, we followed previous work~\cite{DBLP:conf/recsys/YueZKSW22} and set $K$ to  2 for the ML-1M dataset, and 1 for the Steam and Beauty datasets. For other hyperparameters of compared methods, we tuned them via grid search in the ranges provided by their papers or directly took the optimal hyperparameters suggested in their papers ~\cite{DBLP:conf/recsys/YueHZM21,DBLP:conf/recsys/YueZKSW22,DBLP:conf/cikm/SunLWPLOJ19}.

\begin{table*}[]
    \caption{Attack performances of different attack methods \wrt the popularity of the target items, reported in NDCG@10.} 
    \vspace{-8pt}
     \label{tab:attack-popular}
    \resizebox{\linewidth}{!}{
\begin{tabular}{c|cccccc|ccccc|ccccc}
\hline
\multirow{2}{*}{Data} & \multicolumn{1}{c|}{Pop.} & \multicolumn{5}{c|}{Head} & \multicolumn{5}{c|}{Middle} & \multicolumn{5}{c}{Tail} \\ \cline{2-17} 
 & \multicolumn{1}{c|}{Attack} & Clean & RandomAlter & SimAlter & Replace & INFAttack & Clean & RandomAlter & SimAlter & Replace & INFAttack & Clean & RandomAlter & SimAlter & Replace & INFAttack \\ \hline
\multirow{4}{*}{ML-1M} & NARM & 0.202 & 0.205 & 0.206 & 0.352 & \textbf{0.639} & 0.037 & 0.041 & 0.051 & 0.239 & \textbf{0.241} & 0.005 & 0.005 & 0.006 & 0.006 & \textbf{0.008} \\
 & SASRec & 0.217 & 0.257 & 0.320 & 0.651 & \textbf{0.931} & 0.034 & 0.036 & 0.039 & 0.140 & \textbf{0.189} & 0.008 & 0.008 & 0.008 & 0.009 & \textbf{0.121} \\
 & Bert4Rec & 0.201 & 0.211 & 0.236 & 0.270 & \textbf{0.568} & 0.036 & 0.036 & 0.038 & 0.097 & \textbf{0.206} & 0.009 & 0.009 & 0.010 & 0.010 & \textbf{0.014} \\
 & Locker & 0.221 & 0.231 & 0.258 & 0.325 & \textbf{0.538} & 0.037 & 0.038 & 0.043 & 0.142 & \textbf{0.241} & 0.009 & 0.009 & 0.010 & 0.010 & \textbf{0.011} \\ \hline
\multirow{4}{*}{Steam} & NARM & 0.313 & 0.363 & 0.413 & 0.600 & \textbf{0.650} & 0.012 & 0.013 & 0.014 & 0.035 & \textbf{0.075} & 0.000 & 0.000 & 0.000 & 0.000 & \textbf{0.002} \\
 & SASRec & 0.327 & 0.354 & 0.362 & 0.413 & \textbf{0.582} & 0.012 & 0.125 & 0.146 & 0.159 & \textbf{0.167} & 0.000 & 0.000 & 0.000 & 0.000 & \textbf{0.001} \\
 & Bert4Rec & 0.311 & 0.315 & 0.320 & 0.388 & \textbf{0.859} & 0.009 & 0.012 & 0.035 & 0.094 & \textbf{0.137} & 0.000 & 0.000 & 0.000 & 0.000 & \textbf{0.000} \\
 & Locker & 0.318 & 0.319 & 0.385 & 0.428 & \textbf{0.725} & 0.012 & 0.013 & 0.018 & 0.058 & \textbf{0.124} & 0.000 & 0.000 & 0.000 & 0.000 & \textbf{0.001} \\ \hline
\multirow{4}{*}{Beauty} & NARM & 0.261 & 0.319 & 0.335 & 0.805 & \textbf{0.969} & 0.023 & 0.024 & 0.024 & 0.087 & \textbf{0.102} & 0.002 & 0.002 & 0.003 & 0.003 & \textbf{0.005} \\
 & SASRec & 0.246 & 0.417 & 0.524 & 0.549 & \textbf{0.998} & 0.030 & 0.072 & 0.116 & 0.126 & \textbf{0.197} & 0.007 & 0.009 & 0.007 & 0.008 & \textbf{0.011} \\
 & Bert4Rec & 0.260 & 0.269 & 0.285 & 0.469 & \textbf{0.638} & 0.027 & 0.028 & 0.028 & 0.085 & \textbf{0.090} & 0.002 & 0.001 & 0.001 & 0.001 & \textbf{0.002} \\
 & Locker & 0.265 & 0.271 & 0.301 & 0.425 & \textbf{0.628} & 0.031 & 0.031 & 0.038 & 0.096 & \textbf{0.152} & 0.003 & 0.003 & 0.003 & 0.003 & \textbf{0.003} \\ \hline
\multirow{4}{*}{LastFM} & NARM & 0.201 & 0.204 & 0.215 & 0.425 & \textbf{0.528} & 0.028 & 0.029 & 0.042 & 0.145 & \textbf{0.201} & 0.003 & 0.003 & 0.003 & 0.003 & \textbf{0.003} \\
 & SASRec & 0.205 & 0.205 & 0.214 & 0.417 & \textbf{0.528} & 0.031 & 0.032 & 0.045 & 0.124 & \textbf{0.218} & 0.003 & 0.003 & 0.003 & 0.003 & \textbf{0.004} \\
 & Bert4Rec & 0.206 & 0.208 & 0.421 & 0.527 & \textbf{0.585} & 0.036 & 0.036 & 0.042 & 0.128 & \textbf{0.231} & 0.002 & 0.001 & 0.002 & 0.002 & \textbf{0.002} \\
 & Locker & 0.211 & 0.214 & 0.358 & 0.417 & \textbf{0.557} & 0.039 & 0.039 & 0.045 & 0.115 & \textbf{0.225} & 0.004 & 0.004 & 0.005 & 0.004 & \textbf{0.005} \\ \hline
\multirow{4}{*}{ML-20M} & NARM & 0.205 & 0.206 & 0.225 & 0.228 & \textbf{0.631} & 0.032 & 0.034 & 0.036 & 0.137 & \textbf{0.241} & 0.002 & 0.002 & 0.002 & 0.002 & \textbf{0.002} \\
 & SASRec & 0.215 & 0.214 & 0.221 & 0.454 & \textbf{0.759} & 0.035 & 0.034 & 0.037 & 0.138 & \textbf{0.251} & 0.004 & 0.004 & 0.004 & 0.005 & \textbf{0.006} \\
 & Bert4Rec & 0.235 & 0.236 & 0.245 & 0.558 & \textbf{0.668} & 0.038 & 0.035 & 0.041 & 0.138 & \textbf{0.188} & 0.005 & 0.004 & 0.005 & 0.005 & \textbf{0.006} \\
 & Locker & 0.244 & 0.248 & 0.252 & 0.474 & \textbf{0.578} & 0.039 & 0.041 & 0.041 & 0.145 & \textbf{0.221} & 0.005 & 0.005 & 0.005 & 0.006 & \textbf{0.006} \\
 \hline
 % \cline{1-1} \cline{3-17} 
\end{tabular}
}
\end{table*}

\subsection{RQ1: Attack Efficacy}
In this section, we will first examine the overall attack performance of our proposed algorithm INFAttack, and then evaluate its effectiveness in promoting different types of items (popular vs. unpopular items).

 \subsubsection{Overall Attack Performance}
\begin{figure*}
    \centering
    \includegraphics[width=\linewidth]{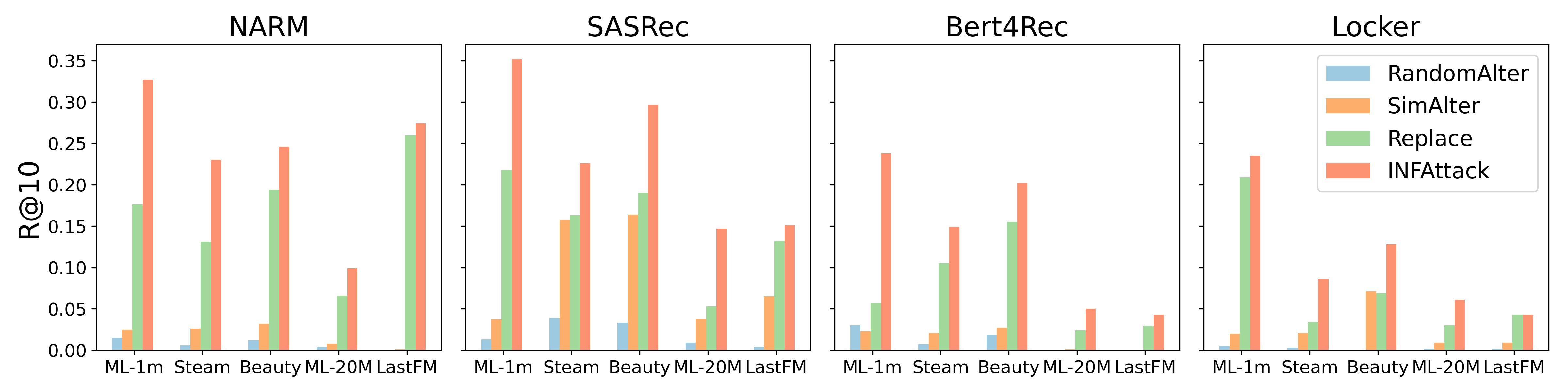}
    \vspace{-20pt}
    \caption{
    The performance change of attack methods compared to the 'Clean' model regarding recommending the target items (\ie attack method - clean model)
    }
     \vspace{-3mm}
    \label{fig:experience-jiagou}
    % \vspace{-2mm}
\end{figure*}

To evaluate the overall attack performance, we randomly sample $15$ items as the target items to promote and report the Recall@10 and NDCG@10 regarding these $15$ target items. Table~\ref{tab:attack-performance} summarizes the results of different attack algorithms for different backbone recommenders. 'Clean' in the table represents the raw results of the backbone model, \ie the results without attacks. To explicitly show the enhancement of attack methods in recommending the target items, we compute the performance improvements of attack methods to Clean (regarding NDCG@10), with the results summarized in Figure~\ref{fig:experience-jiagou}. From the table and figure, we have the following observations:

\begin{itemize}[leftmargin=*]
    \item All attack methods could enhance the recommendations of target items, \ie SimAlter, Replace, and INFAttack have higher NDCG@10 (Recalll@10) about the target items than Clean on each backbone recommender and each dataset. This verifies that all four sequential recommenders are vulnerable to pollution attacks. 

    \item Our method outperforms all baselines in promoting target items on all backbone recommenders and datasets, achieving the best target-item boosting attack. Compared to the best baselines, the averaged relative improvements of INFAttack in NDCG@10 over the datasets are 0.131, 0.133, and 0.126  for  NARM, SASRec, and BERT4Rec, respectively. These results verify the superiority of INFAttack in constructing pollution attacks via the influence function.

    \item The attack efficacy could vary across different backbone recommenders. This can be attributed to that different recommenders are vulnerable to attacks to different degrees. 
    
\end{itemize}

% \textbf{Robustness of different architecture attacks.} Figure(\ref{fig:experience-jiagou}) shows the difference in the robustness of different architectures exhibited by pollution attacks against different recommendation architecture attacks. Through the performance change of N@10 before and after the pollution attack in the figure, we observe that the SASRec and NARM models are more vulnerable to attacks and exhibit higher performance changes. Bert4Rec has the highest average robustness and the worst attack performance. Bert4REC leverages masked training to improve the capture of the global context. In contrast, NARM and SASRec rely on local temporal patterns and are more sensitive.

% \textbf{Item popularity.} 
\subsubsection{Attack Performances \wrt Target Item Popularity} We conduct further experiments to analyze the variation in attack performance with respect to item popularity. We sort items based on their popularity and split them into three groups: Head', which includes the top 20\% most popular items; Tail', which includes the last 20\% of the least popular items; and Middle', which includes all other items. We randomly select five items from each group and study the attack performance of each group. The results are summarized in Table~\ref{tab:attack-popular}. We find that our attack is effective in promoting items with varying levels of popularity. However, for all baselines, attacks aimed at boosting the items in the `Tail' group could become ineffective and may even result in reversed effects, such as decreasing the recommendation of items belonging to the `Tail' group for BERT4Rec on the Beauty dataset.
We hypothesize that the gradient-based method (Replace) may be less effective for unpopular items as the optimal gradient-based attacks are harder to convert into real-world items, considering the difficulty of finding similar items to the target unpopular items, similarly for the similarity-based method (SimAlter). These results further support the superiority of our proposed influence function-based method.

\subsection{RQ2: Impacts of the Attack on Recommendation Performance}
% In addition to promoting target items, maintaining plausible performance is the key to a successful  \cite{nguyen2023poisoning} attack. First, a significant decrease in the overall recommendation performance may cause the attack to fail, since the recommender systems may detect such attacks. In addition, only a high-precision recommender system will have a large user base for promoting the target item, thereby reinforcing the attack target. 
% We thus conduct an experiment to study the recommendation performance between different attack methods and clean models. We summarize the results in Table~\ref{tab:SR-Performance}. According to the table, we can see that our proposed method has the least negative impact on the recommendation performance of the recommendation system in most cases.

In addition to promoting target items, maintaining plausible performance is crucial to the success of a poisoning attack, as noted in~\cite{nguyen2023poisoning}. A significant decrease in the overall recommendation performance can cause the attack to fail, as recommender systems may detect such attacks. Additionally, only a high-precision recommender system can effectively promote the target item to a large user base, reinforcing the attack earnings.
To investigate the impact of different attack methods on recommendation performance, we conduct an experiment and summarize the results in Table~\ref{tab:SR-Performance}. As shown in the table, our proposed method has the least negative impact on the recommendation performance of the system in most cases.

\begin{table}[h]
\caption{The impact of attacks on recommendation performance measured by HR@10 for user-liked items (\ie items truly interacted by the user in the testing set).} 
\vspace{-8pt}
  \label{tab:SR-Performance}
  \resizebox{\linewidth}{!}{
  \begin{tabular}{lcccccl}
    \toprule
      &Method &clean &Random & SimAlter &Replace &INFAttack \\

    \midrule

    \multirow{5}*{NARM} & ML-1M & 0.654& 0.516 & 0.536 &0.612  & \textbf{0.617}\\ 
    \multirow{5}*{} & Steam &0.315  &0.255  &0.268 &\textbf{0.288}  & 0.287\\ 
    \multirow{5}*{} & Beauty &0.262  &0.113  &0.208 &0.224  &\textbf{0.224}\\  
    \multirow{5}*{} & ML-20M &0.689  &0.514  & 0.532&0.618  &\textbf{0.628}\\ 
    \multirow{5}*{} & LastFM &0.312  &0.242  &0.256 &0.261  &\textbf{0.274}\\  \cline{1-7}

\multirow{5}*{SASRec} & ML-1M &0.663 &0.536 	&0.586 	&0.602  &\textbf{0.604}\\ 
    \multirow{5}*{} & Steam &0.378  & 0.265 &0.298 & 0.288 &\textbf{0.323}\\ 
    \multirow{5}*{} & Beauty & 0.265 &0.123  &0.208 &0.224  &\textbf{0.229}\\  
    \multirow{5}*{} & ML-20M &0.714  &0.522  &0.589 & 0.611 &\textbf{0.658}\\ 
    \multirow{5}*{} & LastFM &0.298  &0.185  &0.215 & 0.216 &\textbf{0.234}\\  \cline{1-7}

\multirow{5}*{BERT4Rec} & ML-1M &0.697 &0.434 	&0.455 	&0.535  &\textbf{0.545}\\ 
    \multirow{5}*{} & Steam & 0.401 &0.331  & 0.353& 0.366  &\textbf{0.367}\\ 
    \multirow{5}*{} & Beauty &0.303  &0.235  &0.256 &0.246  &\textbf{0.249}\\  
    \multirow{5}*{} & ML-20M & 0.747 &0.502  &0.615 &0.635  &\textbf{0.685}\\ 
    \multirow{5}*{} & LastFM & 0.302 &0.159  &0.270 &0.271  &\textbf{0.285}\\  \cline{1-7}

\multirow{5}*{Locker} & ML-1M &0.658 & 0.512	&0.552 	&0.548  &\textbf{0.602}\\ 
    \multirow{5}*{} & Steam &0.404  &0.312  & 0.325&0.329  &\textbf{0.365}\\ 
    \multirow{5}*{} & Beauty &0.307  &0.215  &0.254 &0.262  &\textbf{0.270}\\  
    \multirow{5}*{} & ML-20M &0.652  &0.510  &0.531 &0.530  &\textbf{0.611}\\ 
    \multirow{5}*{} & LastFM &0.312  &0.255  &0.265 & 0.267 &\textbf{0.272}\\

  \bottomrule
\end{tabular}
}
\vspace{-4mm}
\end{table}

\begin{figure*}[h]
    \centering
    \includegraphics[width=\linewidth]{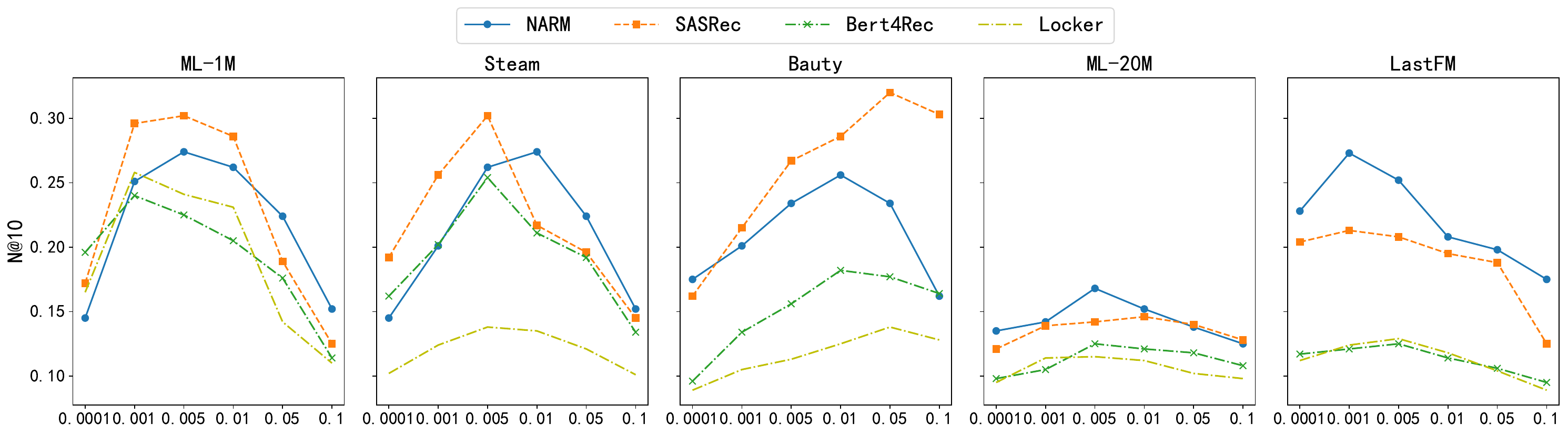}
    \vspace{-20pt}
    \caption{Attack performance of the proposed INFAttack with respect to the damping term (i.e., $\lambda$ in Equation~\eqref{I-theta}), measured by NDCG@10 regarding target items.}
 
    \label{fig:lamba}

\end{figure*}
% \setlength{\abovecaptionskip}{3pt}

% \subsection{RQ3: What role does our proposed influence play in the attack?}

\subsection{RQ3: Importance of the Influence Estimation for the Attack}
% The core of INFAttack is to find the most influential items in promoting the target items via the influence function. We next study the impact of influence function on the attack performance.
% Specifically, we compare INFAttack with a variant INFAttack-NINF that randomly samples injected items to form the polluted sequences.  Table~\ref{tab:INF-effect} reports the comparison results. Compared to INFAttack-NFNF, INFAttack could significantly improve the attack efficacy, \eg, the relative improvements in Recall@10 regarding the target items could reach 300\%. These results show the positive impact of influence estimation on improving attack performance.

The main idea behind INFAttack is to use the influence function to identify the most effective items for promoting the target items. We conducted an experiment to investigate the impact of the influence function on attack performance. Specifically, we compared INFAttack with a variant called INFAttack-NINF, which randomly selects items to create the poisoned sequences. The results are shown in Table~\ref{tab:INF-effect}. As can be seen from the table, INFAttack outperforms INFAttack-NINF by a significant margin. For instance, the relative improvements in Recall@10 for the target items can reach 300\%. These findings demonstrate the effectiveness of using influence estimation to enhance attack performance.

% always performs worse, for example, 

% The key of INFAttack is to combine the influence function to filter the influential additions. In this section, we evaluate the impact of the influence function on attack performance. INFAttack-NINF is an attack with no attack impact loss, i.e. random selection of additional items to append. Table \ref{tab:INF-effect} reports the comparison of INFAttack-NINF with the original INFAttack attack. It reveals that the advantage of using the influential INFAttack is significant. In the best case of Beauty, R@10 improves by about 354\%, while NDCG@10 improves by 484\%. These results emphasize the positive impact of influence on improving attack performance.

\begin{table}[h]
\caption{Attack performance comparison between INFAttack and INFAttack-NINF, where INFAttack-NINF denotes a variant of INFAttack that randomly samples injected items instead of most influential items to perform attacks (R@10).}
\vspace{-8pt}
  \label{tab:INF-effect}
  \resizebox{\linewidth}{!}{
  \begin{tabular}{lcccccl}
    \toprule
     & Dataset & ML-1M &Steam &Beauty & LastFM & Ml-20M \\

    \midrule
    \multirow{2}*{NARM} & INFAttack-NINF  & 0.142	& 0.120 & 0.130& 0.148 &0.158\\ 
    \multirow{2}*{} & INFAttack  & 0.467 & 0.352  & 0.372 &0.421&0.256\\  \cline{1-7}

    \multirow{2}*{SASRec} & INFAttack-NINF  & 0.137  & 0.125		& 0.135 &0.164&0.173\\ 
    \multirow{2}*{} & INFAttack  & 0.487  & 0.355  & 0.432&0.315&0.320\\  \cline{1-7}
    
    \multirow{2}*{BERT4Rec} & INFAttack-NINF  & 0.152	 & 0.132  & 0.152 &0.155 &0.157\\ 
    \multirow{2}*{} & INFAttack & 0.364  & 0.278	 & 0.345  &0.205 &0.200\\  \cline{1-7}
    
      \multirow{2}*{Locker} & INFAttack-NINF  & 0.152	& 0.132  & 0.152 &0.157 &0.150\\ 
    \multirow{2}*{} & INFAttack & 0.364  & 0.278	 & 0.345 &0.200 &0.211 \\ 
    
  \bottomrule
\end{tabular}
}
\end{table} 

\subsection{RQ4: The Impact of Hyper-parameters}

% In this section, we study the effects of two hyper-parameters ($K$ to control the number of inject items and the damping term $\lambda$) on the attack and recommendation performance.

In this subsection, we investigate the impact of two hyper-parameters, namely $K$  (to control the number of injected items) and the damping term $\lambda$ in Equation~\eqref{I-theta}, on the attack and recommendation performance.

\begin{figure}[h]
    \centering
    \includegraphics[width=\linewidth]{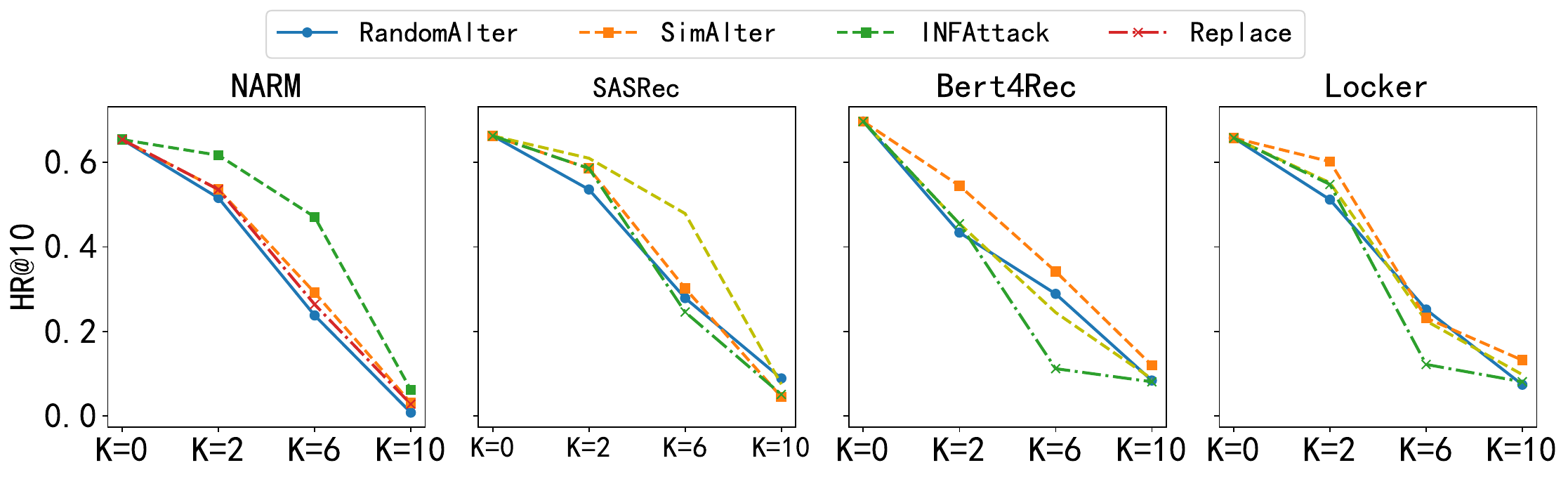}
    \vspace{-20pt}
    \caption{ The attack performance of the attack methods with respect to the number of injected items (\ie K), evaluated using the NDCG@10 metric for target items.}
    \vspace{-2mm}
    \label{fig:experience6}
   
\end{figure}

\begin{figure}[h]
    \centering
    \includegraphics[width=\linewidth]{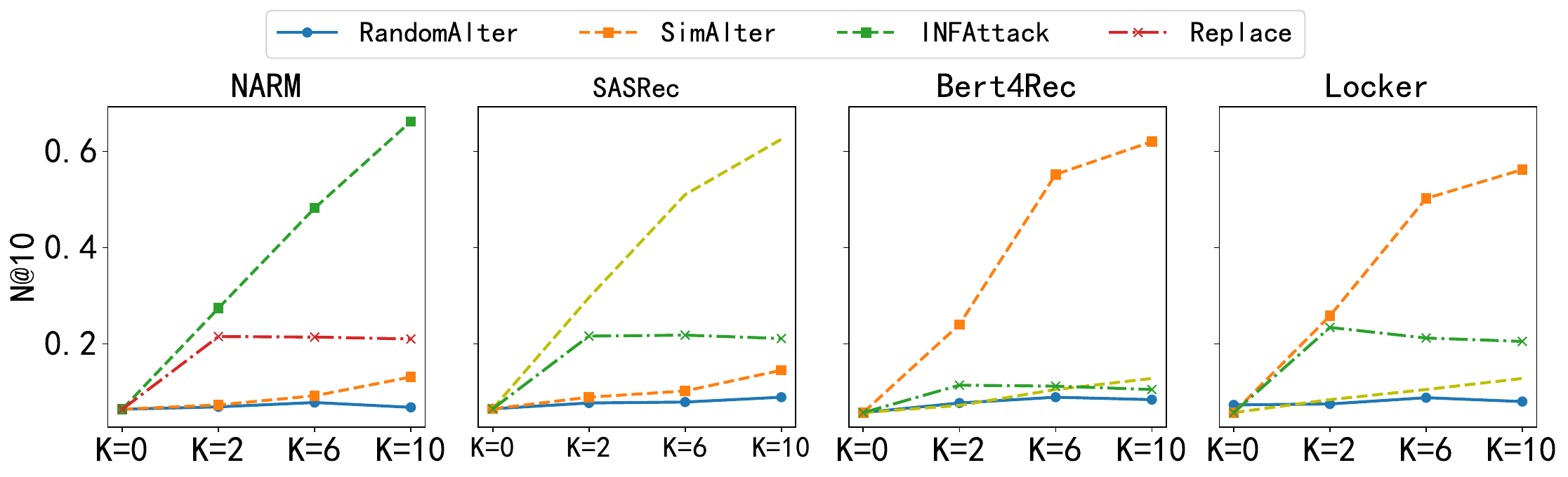}
    \caption{The recommendation performance of the attack methods with respect to the number of injected items (\ie K), evaluated using the HR@10 metric for user-liked items.}
    \vspace{-3mm}
    \label{fig:experience6b}
    % \vspace{-2mm} 
\end{figure}

\noindent\textbf{The effects of $K$.} We first investigate the impact of the additional items injected into the user interaction profile ($K$) on the attack method's performance. We use four values of $K$ (0, 2, 6, and 10) on the ML-1M dataset (results are consistent with other datasets) to examine how the number of injection items affects the attack's properties, including attack performance (Figure~\ref{fig:experience6}) and recommendation performance~(Figure~\ref{fig:experience6b}). Figure \ref{fig:experience6} shows that increasing the number of injections improves the attack performance for all methods. However, adding too many injection items can introduce extra noise and significantly decrease the recommendation performance of the recommendation system, as shown in Figure~\ref{fig:experience6b}. Therefore, to balance the attack effect and the performance impact on the recommendation system, we choose the $K$ value ($K=2$) that has the least impact on the recommendation performance of the recommendation system in former experiments. Besides, we observed that as K increases, our method's curve for recommendation performance becomes flatter, while the curve for attack performance becomes steeper. This indicates that our method can achieve a greater attack effect with a smaller cost in recommendation performance. It also means that within the allowable range of recommendation performance loss, we could insert more items to enhance the attack.

% The size of the number $K$ of additional items injected into the user interaction profile is an important factor for the attack method. In this section, we will use three $K$ values in the Beauty dataset (results consistent with other datasets) to examine its impact on the attack $K = {2, 6, 10}$ to explore items that change the injection profile The number of effects attack properties. It can be seen from Figure(\ref{fig:experience6}) that when the number of injections increases, the attack performance of this attack method will continue to improve. 
% At the same time, we explored the influence of the $K$ value on the recommendation accuracy of the recommendation system. As shown in Figure(\ref{fig:experience6b}), using too many injection items will lead to a significant decline in the recommendation performance of the recommendation system because they will introduce extra noise. Based on these, in order to balance the attack effect and the performance impact on the recommendation system, we choose the $K$ value ($K=2$) that has the least impact on the recommendation performance of the recommendation system.

\noindent\textbf{The effects of $\lambda$.}  
We next investigate how the damping term $\lambda$ in Equation~\ref{I-theta}, which is used to make the Heassin matrix (see Equation~\eqref{I-theta}) invertible, affects the attack performance. We conduct an experiment by varying the parameter in the range of \{0.0001, 0.001, 0.005, 0.01, 0.05, 0.1\}. The results are shown in Figure~\ref{fig:lamba}.  We find the attack performance of our INFAttack increases first and then decreases as $\lambda$ increases. This is because a small $\lambda$ cannot make sure the Hessian matrix ($H_{\theta}+\lambda I$) is invertible, while a large $\lambda$ could introduce significant noise. This verifies the necessity of using the damping term, however, we should set it with an appropriate value. Meanwhile, we could find that optimal $\lambda$ in most cases could be very close, and the most common optimal $\lambda$ is 0.005 or 0.01 for all datasets and backbones. This means that we may not need to spend too much effort on tuning this hyper-parameter for different datasets and backbones.

\section{Conclusion}

In this work, we study the effectiveness of influence function-based profile pollution attacks on sequential recommender systems and introduce a new influence function-based approach called INFAttack. We conduct extensive experiments using five benchmark datasets. The results clearly show that INFAtack outperforms baseline methods in generating effective attacks. However, INFAtack has some limitations, and our future research directions include wider application of influence functions in attack scenarios. For example, we plan to identify the users most influential to potential attacks, explore more efficient ways to estimate influence function and develop a more comprehensive framework for their exploitation. Additionally, we intend to create robust training and defense strategies to enhance the system's ability to withstand proposed attacks.


\bibliographystyle{ACM-Reference-Format}
\bibliography{sample-base}
  
\end{document}